\documentclass[onecolumn,prx,amsmath,amssymb,floatfix,longbibliography,notitlepage,nofootinbib]{revtex4-1} %rmp
\usepackage{calrsfs}
\DeclareMathAlphabet{\pazocal}{OMS}{zplm}{m}{n}
\usepackage{graphicx}
\usepackage{float}
\usepackage{amssymb,amsthm,amsfonts,amstext,amsmath}
\usepackage{url}
\usepackage{xcolor}
\usepackage{tcolorbox}
\usepackage{enumitem}
\usepackage{tikz}
\usepackage{textpos}

\def\be{\begin{equation}}
\def\ee{\end{equation}}
\def\bea{\begin{eqnarray}}
\def\eea{\end{eqnarray}}
\def\bma{\begin{mathletters}}
\def\ema{\end{mathletters}}

\def\P{{\cal P}}

\def\q0{\underline{0}}

\def\H{{\cal H}}

\def\T{{\cal T}}
\def\P{{\cal P}}
\def\S{{\cal S}}
\def\C{{\mathbb C}}
\def\id{{\mathbb I}}
\def\E{{\cal E}}

\def\M{{\cal M}}
\def\H{{\cal H}}
\def\C{{\cal C}}

\def\R{\mathbb{R}}

\def\tr{\mbox{tr}}

\def\one{\leavevmode\hbox{\small1\normalsize\kern-.33em1}}

\def\bra#1{\langle#1|} \def\ket#1{|#1\rangle}

\def\proj#1{\ket{#1}\!\bra{#1}}

\def\G{{\cal G}}
\def\score{\gamma}

\def\del#1{{\color{red} %#1
}}
\renewcommand{\emph}[1]{{#1}}

\newtheorem{assumption}{Assumption}
\def\id{{\mathbb I}}

\begin{document}

\title{Quantum Preparation Games}
\author{M. Weilenmann$^1$, E. A.  Aguilar$^{1,2}$,  M. Navascu\'es$^1$}

\affiliation{$^1$Institute for Quantum Optics and Quantum Information (IQOQI) Vienna\\ Austrian Academy of Sciences\\$^2$AIT Austrian Institute of Technology GmbH, 1210 Vienna, Austria}

\begin{abstract}
	A preparation game is a task whereby a player sequentially sends a number of quantum states to a referee, who probes each of them and announces the measurement result. Many experimental tasks in quantum information, such as entanglement quantification or magic state detection, can be cast as preparation games. In this paper, we introduce general methods to design $n$-round preparation games, with tight bounds on the performance achievable by players with arbitrarily constrained preparation devices. We illustrate our results by devising new adaptive measurement protocols for entanglement detection and quantification. Surprisingly, we find that the standard procedure in entanglement detection, namely, estimating n times the average value of a given entanglement witness, is in general suboptimal for detecting the entanglement of a specific quantum state. On the contrary, there exist $n$-round experimental scenarios where detecting the entanglement of a known state optimally requires adaptive measurement schemes.
\end{abstract}

\maketitle

\section{Introduction} 

Certain tasks in quantum communication can only be conducted when all the parties involved share a quantum state with a specific property. For instance, two parties with access to a public communication channel must share an entangled quantum state in order to generate a secret key \cite{QKD}. If the same two parties wished to carry out a qudit teleportation experiment, then they would need to share a quantum state with an entanglement fraction beyond $1/d$ \cite{teleportation}. 
More generally, when only restricted quantum operations are permitted, specific types of quantum states become instrumental for completing certain information processing tasks. This is usually formalized in terms of resource theories~\cite{resource}. Some resources, like entanglement, constitute the basis of quantum communication. Others, such as magic states, are required to carry out quantum computations \cite{magic}. Certifying and quantifying the presence of resource states with a minimum number of experiments is the holy grail of entanglement \cite{reviewEnt} and magic state detection \cite{magic}. 

Beyond the problem of characterizing resourceful states mathematically, the experimental detection and quantification of resource states is further complicated by the lack of a general theory to devise efficient measurement protocols. Such protocols would allow one to decide, at minimum experimental cost, whether a source is capable of producing resourceful states. Developing such methods is particularly important for high dimensional systems where full tomography is infeasible or in cases where the resource states to be detected are restricted to a small (convex) subset of the state space, which renders tomography excessive.

General results on the optimal discrimination between different sets of states in the asymptotic regime~\cite{renyi} suggest that the optimal measurement protocol usually involves collective measurements over many copies of the states of interest, and thus would require a quantum memory for its implementation. This contrasts with the measurement scenario encountered in many experimental setups: the lack of a quantum memory often forces an experimentalist to measure each of the prepared states as soon as they arrive at the lab. In this case it is natural to consider a setting where subsequent measurements can depend on previous measurement outcomes, in which case the experimentalist is said to follow an adaptive strategy. Perhaps due to their perceived complexity, the topic of identifying optimal adaptive measurement strategies has been largely overlooked in quantum information theory.

In this paper, we propose the framework of quantum preparation games to reason about the detection and quantification of resource states in this adaptive setting. These are games wherein a player will attempt to prepare some resource which the referee will measure and subsequently assign a score to. We prove a number of general results on preparation games, including the efficient computation of the maximum average score achievable by various types of state preparation strategies. Our results furthermore allow us to optimise over the most general measurement strategies one can follow with only a finite set of measurements, which we term Maxwell demon games. Due to limited computational resources, full optimisations over Maxwell demon games are restricted to scenarios with only $n\approx 3, 4$ rounds. For higher round numbers, say, $n\simeq 20$, we propose a heuristic, based on coordinate descent, to carry these optimisations out approximately. More specifically, the outcome of the heuristic is (in general) a sub-optimal preparation game that nonetheless satisfies all the optimization constraints. In addition, we show how to devise arbitrarily complex preparation games through game composition, and yet another heuristic inspired by gradient descent. We illustrate all our techniques with examples from entanglement certification and quantification and highlight the benefit of adaptive measurement strategies in various ways. In this regard, in contradiction to standard practice in entanglement detection, we find that the optimal $n$-round measurement protocol to detect the entanglement of a single, known quantum state does not consist in estimating $n$ times the value of a given (optimised) entanglement witness. On the contrary, there exist adaptive measurement schemes that supersede any non-adaptive protocol for this task.

\section{Quantum Preparation games for resource certification and quantification}
\label{sec:def}

Consider the following task: a source is distributing multipartite quantum states, $\rho_{1,...,m}$, among a number of separate parties who wish to quantify how entangled those states are. To this effect, the parties sequentially probe a number of $m$-partite states prepared by the source. Depending on the results of each experiment, they decide how to probe the next state. After a fixed number of rounds, the parties estimate the entanglement of the probed states. They naturally seek an estimate that lower bounds the actual entanglement content of the states produced during the experiment with high probability. Most importantly, if the source is unable to produce entangled states, the protocol should certify this with high probability.

Experimental scenarios whereby a source (or player) sequentially prepares quantum states that are subject to adaptive measurements (by some party or set of parties that we collectively call the referee) are quite common in quantum information. Besides entanglement detection, they are also found in magic state certification \cite{magic}, and, more generally, in the certification and quantification of any quantum state resource \cite{resource}. The common features of these apparently disparate quantum information processing tasks motivate the definition of \emph{quantum preparation games}.

\smallskip

\begin{figure}
	\begin{minipage}[t]{0.96\linewidth}
		\includegraphics[trim=1cm 8cm 0.5cm 5cm,clip,width=1\textwidth]{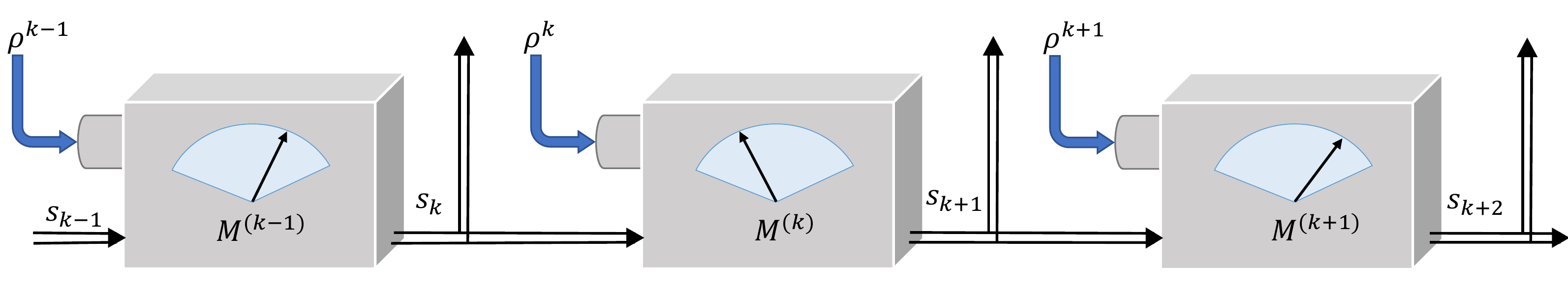}
	\end{minipage}
	\caption{Quantum preparation game from the referee's perspective. In each round $k$ of a preparation game, the referee (measurement box) receives a quantum state $\rho^k$ from the player. The referee's measurement $M^{(k)}$ will depend on the current game configuration $s_k$, which is determined by the measurement outcome of the previous round. In the same way, the outcome $s_{k+1}$ of round $k$ will determine the POVMs to be used in round $k+1$. Recall that the player can tailor the states $\rho^k$ to the measurements to be performed in round $k$, since they have access to the (public) game configuration $s_k$, shown with the upward line leaving the measurement apparatus. }
	\label{fig:preparationgameround}
\end{figure}

\begin{tcolorbox}[colback=black!2!white,colframe=blue!35!black,title= Box~1: Quantum Preparation Game]
\textbf{Game Variables}

\textit{Number of Rounds:} $n$.

\textit{Game Configuration:} There is a unique initial game configuration $S_1 = \{\emptyset\}$. At every round $k$, there is a set of allowed configurations $S_k = \{s^k_1, s^k_2,\ldots\}$. After $n$ rounds, the game ends in one of the final configurations $s \in S_{n+1}$.

\textit{Measurement Operators:} For every game configuration $s\in S_k$, there are POVMs $\{M^{(k)}_{s'|s} : s'\in S_{k+1}\}$.

\textit{Scoring Function:} A (non-deterministic) function $g: S_{n+1} \rightarrow \mathbb{R}$.

\smallskip

The game variables, i.e., round number, possible configurations, POVMs and scoring rule, are publicly announced before the game starts.

\textbf{Measurement Round Rules}

At the beginning of round $k$, the current game configuration $s \in S_k$ is known to the player. The player prepares a state $\rho^k$ according to their preparation strategy $\mathcal{P}$, and sends it to the referee.

The referee measures the quantum state $\rho^k$ with the POVM $\{M^{(k)}_{s'|s} : s'\in S_{k+1}\}$.

The referee publicly announces the outcome $s'$ of this measurement, which becomes the game configuration for the next round.

\textbf{Scoring}

After the $n^{\text{th}}$ round, the player receives a score $g(s)$, where $s\in S_{n+1}$ is the final configuration. 

\smallskip

See also Fig.~\ref{fig:preparationgameround} for a pictorial representation of the  procedure followed in round $k$. 

\end{tcolorbox}

A preparation game $G$ is thus fully defined by the triple $(S, M, g)$, where $S$ denotes the sequence of game configuration sets $(S_k)_{k=1}^{n+1}$; and $M$, the set of POVMs $M \equiv \{M^{(k)}_{s'|s}:s'\in S_{k+1}, s\in S_{k}\}_{k=1}^n$. In principle, the Hilbert space where the state prepared in round $k$ lives could depend on $k$ and on the current game configuration $s_k\in S_k$. For simplicity, though, we will assume that all prepared states act on the same Hilbert space, $\H$. In many practical situations, the actual, physical measurements conducted by the referee in round $k$ will have outcomes in $O$, with $|O|<|S_k|$. The new game configuration $s'\in S_{k+1}$ is thus decided by the referee through some non-deterministic function of the current game configuration $s$ and the `physical' measurement outcome $o\in O$. The definition of the game POVM $\{M^{(k)}_{s'|s}\}_{s'}$ encompasses this classical processing of the physical measurement outcomes.

The expected score of a player with preparation strategy $\mathcal{P}$ is
\be
G(\P)\equiv\sum_{s\in S_{n+1}} p(s|\P, G)\langle g(s)\rangle.
\label{eq:gameobjective}
\ee
In the equation, $p(s|\P, G)$ denotes the probability that, conditioned on the player using a preparation strategy $\P$ in the game $G$, the final game configuration is $s$. For the sake of clarity, we will sometimes refer to the set of possible final configurations as $\bar{S}$ instead of $S_{n+1}$.

\begin{figure}
	\begin{minipage}[t]{0.96\linewidth}
		\includegraphics[trim=1cm 6.8cm 0.5cm 5.8cm,clip,width=1\textwidth]{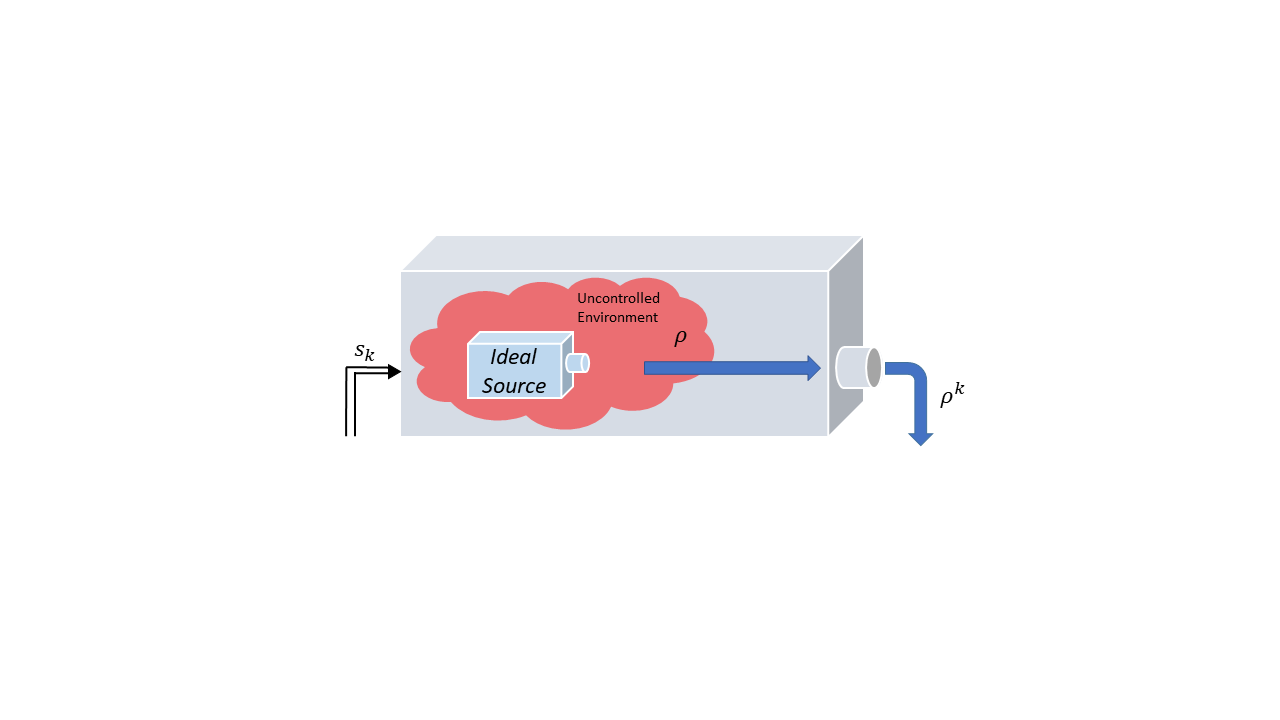}
	\end{minipage}
	\caption{Finitely correlated strategies.	
	Suppose that a player owns a device which allows them to prepare and distribute a quantum state to the referee. Unfortunately, at each experimental preparation the player's device interacts with an environment $A$. Explicitly, if the player activates their device, then the referee receives the state
	$
		\tr_{A}\left[ \sum_i K_i \rho K_i^\dagger\right],$
		where $\rho$ is the current state of the environment and $K_i:\H_A\to \H_A\otimes\H$ are the Kraus operators which evolve the environment and prepare the state that the referee receives. Since the same environment is interacting with each prepared state, the states that the referee receives in different rounds are likely correlated.
}
	\label{fig:player}
\end{figure}

In this paper we consider players who aim to maximise their expected score over all preparation strategies $\P$ that are accessible to them, in order to convince the referee of their ability to prepare a desired resource.
Intuitively, a preparation strategy is the policy that a player follows to decide, in each round, which quantum state to prepare. Since the player has access to the referee's current game configuration, the player's state preparation can depend on this. The simplest preparation strategy, however, consists in preparing independent and identically distributed (i.i.d.) copies of the same state $\rho$. We call such preparation schemes \emph{i.i.d.\ strategies} and denote them as $\rho^{\otimes n}$. A natural extension of i.i.d.\ strategies, which we call finitely correlated strategies~\cite{wernerfinitely}, follows when we consider interactions with an uncontrolled environment, see Figure~\ref{fig:player}. I.i.d.\ and finitely correlated strategies can be extended to scenarios where the preparation  depends on the round number $k$. The mathematical study of these strategies is so similar to that of their round-independent counterparts, that we will not consider such extensions in this article.

Instead, we will analyze more general scenarios, where the player is limited to preparing multipartite states belonging to a specific class $\C$, e.g.\ separable states. 
In this case, given $\rho,\sigma\in\C\cap B(\H)^{\otimes k}$, a player can also generate the state $p\rho +(1-p)\sigma$ for any $p \in [0,1]$, just by preparing $\rho$ with probability $p$ and $\sigma$ otherwise. Thus, we can always assume $\C\cap B(\H)^{\otimes k}$ to be convex for all $k$. The preparation strategies of such a player will be assumed fully general, e.g., the state preparation in round $k$ can depend on $k$, or on the current game configuration $s_k$. We call such strategies \emph{$\C$-constrained}.

\subsection{Computing the average score of a preparation game}

Even for i.i.d.\ strategies, a brute-force computation of the average game score  would require adding up a number of terms that is exponential in the number of rounds. In the following we introduce a method to efficiently compute the average game scores for various types of player strategies.

Let $G=(S, M, g)$ be a preparation game with $M \equiv \{M^{(k)}_{s'|s}:s'\in S_{k+1}, s\in S_{k}\}_{k=1}^n$, and let $\C$ be a set of quantum states. In principle, a $\C$-constrained player could exploit correlations between the states they prepare in different rounds to increase their average score when playing $G$. They could, for instance, prepare a bipartite state $\rho^{12}\in\C$; send part $1$ to the referee in round $1$ and, depending on the referee's measurement outcome $s_2$, send part $2$, perhaps after acting on it with a completely positive map depending on $s_2$. However, the player would be in exactly the same situation if, instead, they sent state $\rho^1=\tr_2(\rho^{12})$ in round $1$ and state $\rho^{2}_{s_2}\propto\tr_1\left[(M_{s_2|\emptyset}\otimes \id_2)\rho^{12}\right]$ in round $2$. There is a problem, though: the above is only a $\C$-constrained preparation strategy provided that $\rho^{2}_{s_2}\in \C$. This motivates us to adopt the following assumption.

\begin{assumption}
\label{closure}
The set of (in principle, multipartite) states ${\cal C}$ is closed under arbitrary postselections with the class of measurements conducted by the referee.
\end{assumption}
\noindent This assumption holds for general measurements when $\C$ is the set of fully separable quantum states or the set of states with entanglement dimension \cite{Terhal1999} at most $D$ (for any $D > 1$). It also holds when $\C$ is the set of non-magic states and the referee is limited to conducting convex combinations of sequential Pauli measurements~\cite{Veitch_2014}. More generally, the assumption is satisfied when, for some convex resource theory \cite{resource}, $\C$ is the set of resource-free states; and the measurements of the referee are resource-free. The assumption is furthermore met when the player does not have a quantum memory.

Under Assumption~\ref{closure}, the player's optimal $\C$-constrained strategy consists in preparing a state $\rho^k_{s_k}\in \C$ in each round $k$, depending on  $k$ and the current game configuration $s_k$. Now, define $\mu^{(k)}_{s}$ as the maximum average score achieved by a player, conditioned on $s$ being the configuration in round $k$. Then $\mu^{(k)}_{s}$ satisfies
\begin{align}
&\mu^{(n)}_s = \max_{\rho\in \C}\sum_{\bar{s}\in \bar{S}}\tr[M^{(n)}_{\bar{s}|s}\rho]\langle g(\bar{s})\rangle,\nonumber\\
&\mu^{(k)}_s =\max_{\rho\in \C}\sum_{s'}\tr[M^{(k)}_{s'|s}\rho]\mu^{(k+1)}_{s'}.
\label{induction}
\end{align}

\noindent These two relations allow us to inductively compute the maximum average score achievable via $\C$-constrained strategies, $\mu^{(1)}_\emptyset$. Note that, if the optimizations above were carried out over a larger set of states $\C'\supset \C$, the end result would be an upper bound on the achievable maximum score. This feature will be handy when $\C$ is the set of separable states, since the latter is difficult to characterize exactly \cite{NP_sep, strong_NP_sep}. In either case, the computational resources to conduct the computation above scale as $O\left(\sum_k|S_k||S_{k+1}|\right)$.

Equation \eqref{induction} can also be used to compute the average score of an i.i.d.\ preparation strategy $\rho^{\otimes n}$. In that case, $\C=\{\rho\}$, and the maximization over $\C$ is trivial. Similarly, an adaptation of \eqref{induction} allows us to efficiently compute the average score of finitely correlated strategies, for the details we refer to Appendix~\ref{app:fincorr}.

\subsection{Optimizing preparation games}
\label{optim_sec}
\label{sec:optim}

Various tasks in quantum information -- including entanglement detection -- have the following structure: given two sets of preparation strategies ${\S}, {\S}'$ and a score function $g$, we want to find a game $G=(S,M,g)$ that separates these two sets, i.e., a game such that $G(\P)\leq \delta$, for all $\P\in \S$, and $G(\P)>\delta$ for all $\P\in \S'$. In some cases, we are interested to search for games where the POVMs conducted by the referee belong to a given (convex) class ${\cal M}$. This class represents the experimental limitations affecting the referee, such as space-like separation or the unavailability of a given resource.

Finding a preparation game satisfying the above constraints can be regarded as an optimization problem over the set of quantum preparation games. Consider a set ${\pazocal M}$ of adaptive measurement protocols of the form $M\equiv\{M^{(k)}_{s'|s}:s'\in S_{k+1}, s\in S_k\}$, a selection of preparation games $\{G^i_M=(S, M, g^i)\}_{i=1}^{r}$ and sets of preparation strategies $\{\S_i\}_{i=1}^r$. A general optimization over the set of quantum preparation games is a problem of the form 
\begin{align}
&\min_{M\in{\pazocal M},v} f(v)\nonumber\\
\mbox{s.t. }&G^i_M(\P)\leq v_i, \ \forall\P\in\S_i,i=1,...,r,\nonumber\\
&A\cdot v\leq b,
\label{optim}
\end{align}
\noindent where $A$, $b$ are a $t\times r$ matrix and a vector of length $t$, respectively, and $f(v)$ is assumed to be convex on the vector $v\in \R^r$.

In this paper, we consider i.i.d., finitely correlated (with known or unknown environment state) and $\C$-constrained preparation strategies. The latter class also covers scenarios where a player wishes to play an i.i.d.\ strategy with an imperfect preparation device. Calling $\rho$ the ideally prepared state, one can model this contingency by assuming that, at every use, the preparation device (adversarially) produces a quantum state $\rho'$ such that $\|\rho-\rho'\|_1\leq \epsilon$. If, independently of the exact states prepared by the noisy or malfunctioning device, we wish the average score $g^i$ to lie below some value $v_i$, then the corresponding constraint is
\be
G^i_M(\P)\leq v_i, \forall\P\in \cal{E}_{\epsilon},
\label{errors_prep}
\ee
\noindent where $\cal{E}_{\epsilon}$ is the set of $\epsilon$-constrained preparation strategies, producing states in $\{ \rho' : \rho'\geq 0,\tr(\rho')=1, \| \rho' - \rho \|_1 \leq \epsilon \}$.

The main technical difficulty in solving problem \eqref{optim} lies in expressing conditions of the form 
\be
G_M(\P)\leq v,\forall\P\in\S
\label{upper_score}
\ee
\noindent in a convex (and tractable) way. This will, in turn, depend on which type of measurement protocols we wish to optimize over. We consider families of measurement strategies $M$ such that the matrix
\be
\sum_{s_2,...,s_{n+1}}\langle g(s_{n+1})\rangle\bigotimes_{k=1}^n (M^k_{s_{k+1}|s_{k}}\otimes\ket{s_{k+1}})
\label{linear_object}
\ee
\noindent depends affinely on the optimization variables of the problem. For  $\S=\{\P\}$, condition \eqref{upper_score} then amounts to enforcing an affine constraint on the optimization variables defining the referee's measurement strategy. For finitely correlated strategies, we describe in Appendix~\ref{app:fincorrconstr} how to phrase \eqref{upper_score} as a convex constraint.

For $\C$-constrained strategies, the way to express \eqref{upper_score} as a convex constraint depends more intricately on the class of measurements we aim to optimize over. Let us first consider preparation games with $n=1$ round, where we allow the referee to conduct any $|\bar{S}|$-outcome measurement from the convex set $\M$. Let $\S$ represent the set of all $\C$-constrained preparation strategies, for some convex set of states $\C$. Then, condition \eqref{upper_score} is equivalent to
\be
v \id -\sum_{s\in \bar{S}}M^{(1)}_{s|\emptyset}\langle g(s)\rangle\in \C^*.
\label{dual_cond}
\ee
\noindent 
Note that, if we replace $\C^*$ in \eqref{dual_cond} by a subset thereof, relation \eqref{upper_score} is still implied. In that case, however, there may be values of $v$ for which relation \eqref{upper_score} holds, but not eq.~\eqref{dual_cond}. As we will see later, this observation allows us to devise sound entanglement detection protocols, in spite of the fact that the dual of the set of separable states is difficult to pin down \cite{NP_sep, strong_NP_sep}.

Next, we consider a particularly important family of multi-round measurement schemes, which we call \emph{Maxwell demon games}. In a Maxwell demon game, the referee's physical measurements in each round $k$ are taken from a discrete set ${\cal M}(k)$. Namely, for each $k$, there exist sets of natural numbers $A_k, X_k$ and fixed POVMs $\{(N^{(k)}_{a|x}:a\in A_k):x\in X_k\}\subset B(\H)$. The configuration space at stage $k$ corresponds to the complete history of physical inputs $x_1, \ldots, x_{k-1}$ and outputs $a_1, \ldots, a_{k-1}$, i.e., $s_k=(a_1,x_1,...,a_{k-1},x_{k-1})$, where $s_1=\emptyset$. Note that the cardinality of $S_k$ grows exponentially with $k$. In order to decide which physical setting $x_k$ must be measured in round $k$, the referee receives advice from a Maxwell demon. The demon, who holds an arbitrarily high computational power and recalls the whole history of inputs and outputs, samples $x_k$ from a distribution $P_k(x_k|s_k)$. The final score of the game $\score\in \G$ is also chosen by the demon, through the distribution $P(\score|s_{n+1})$. A Maxwell demon game is the most general preparation game that a referee can run, under the reasonable assumption that the set of experimentally available measurement settings is finite. 

Let us consider
\be
P(x_1,...,x_n, \score|a_0, a_1,...,a_n)=P(\score|s_{n+1}) \prod_{k=1}^n P_k(x_k|s_k),
\label{reduction}
\ee
where $a_0=\emptyset$. Define $(y_0,...,y_n)\equiv (x_1,...,x_n,\score)$. As shown in~\cite{costantino}, a collection of normalized distributions $P(y_0,...,y_{n}|a_0,...,a_n)$ admits a decomposition of the form (\ref{reduction}) iff the \emph{no-signalling-to-the-past} conditions
\be
\sum_{y_{k+1},...,y_{n}}P(y_0,...,y_{n}|a_0,...,a_n)=P(y_0,...,y_k|a_0,...,a_k) \quad \forall k
\label{NSP}
\ee
\noindent hold. For completeness, the reader can find a proof in Appendix~\ref{app:maxwelldemon}. We can thus characterize general Maxwell demon games through finitely many linear constraints on $P(x_1,...,x_n, \score|a_0, a_1,...,a_n)$.

For Maxwell demon games, the matrix \eqref{linear_object} depends linearly on the optimization variables $P(x_1,...,x_n, \score|a_0, a_1,...,a_n)$. Hence, we can express condition \eqref{upper_score} as a tractable convex constraint whenever $\S$ corresponds to an i.i.d.\ strategy, or a finitely correlated strategy with an unknown initial environment state, as described above. Enforcing \eqref{upper_score} for $\C$-constrained strategies requires regarding the quantities $\{\mu_s^{(n)}\}_s$ in eq.~\eqref{induction} as optimization variables, related to $P$ and to each other through a dualized version of the conditions~\eqref{induction}. The reader can find a full explanation in Appendix~\ref{app:maxwelldemon}.

Finally, we consider the set of adaptive measurement schemes with fixed POVM elements $\{M^{(j)}_{s'|s}:j\not=k\}$ and variable $\{M^{(k)}_{s'|s}\}\subset \M$, for some tractable convex set of measurements $\M$. As in the two previous cases, the matrix (\ref{linear_object}) is linear in the optimization variables $\{M^{(k)}_{s'|s}\}$, so \eqref{upper_score} can be expressed in a tractable, convex form for sets of finitely-many strategies and finitely correlated strategies with unknown environment. 
Similarly to the case of Maxwell demon games, enforcing \eqref{upper_score} for $\C$-constrained strategies requires promoting $\{\mu^{(j)}_s:j\leq k\}$ to optimization variables (see Appendix). 

Via coordinate descent, this observation allows us to conduct optimizations (\ref{optim}) over the set of all adaptive schemes with a fixed game configuration structure $(S_j)_{j=1}^{n+1}$. Consider, indeed, the following method.

\begin{tcolorbox}[colback=black!2!white,colframe=blue!35!black,title= Box~2: A heuristic for general optimizations over preparation games]
\begin{enumerate}
\item
Starting point: a natural number $L$, an optimization problem of the form (\ref{optim}), a sequence of sets of game configurations $S=(S_j)_{j=1}^{n+1}$, a measurement scheme $M=\{M^{(j)}_{s_{j+1}|s_j}:s_{j+1}, s_j\}_j$ such that $G^i_M(\P)\leq v_i,\forall \P\in\S_i$, for $i=1,...,r$, with $A\cdot v\leq b$.
\item
Set $l=1$.
\item
Choose an index $k\in\{1,...,n\}$ and, using the techniques explained in Appendix~\ref{app:implementationfincorr} (eq.\ \ref{constOpt}), minimize the objective value of \eqref{optim} over measurement schemes $\tilde{M}$ with $\tilde{M}^{(j)}_{s_{j+1}|s_j}=M^{(j)}_{s_{j+1}|s_j}$, for all $s_j\in S_j,s_{j+1}\in S_{j+1},j\not=k$, subject to the optimization constraints. Call $f^\star$ the objective value of the optimal measurement scheme $M^\star$.
\item
$M\leftarrow M^\star$, $l\leftarrow l+1$. If $l\geq L$, return $M$ and $f^\star$ and stop. Otherwise, go to step $3$.
\end{enumerate}
\end{tcolorbox}

\noindent With this algorithm, at each iteration, the objective value $f(v)$ in problem \eqref{optim} can either decrease or stay the same: The hope is that it returns a small enough value $f^\star$ after a moderate number $L$ of iterations. In Appendix~\ref{app:implementationfincorr} the reader can find a successful application of this heuristic to devise $20$-round quantum preparation games.

The main drawback of this algorithm is that it is very sensitive to the initial choice of POVMs, so it generally requires several random initializations to achieve a reasonably good value of the objective function. It is therefore suitable for optimizations of $n\approx 50$ round measurement schemes.  Optimizations over, say, $n=1000$ round games risk getting stuck in a bad local minimum. 

To address this issue, we provide two additional methods for the design of large-$n$ quantum preparation games below.

\subsection{Large-round preparation games from composition}

The simplest way to construct preparation games with arbitrary round number consists in playing several preparation games, one after another. 
Consider thus a game where, in each round and depending on the current game configuration, the referee chooses a preparation game. Depending on the outcome, the referee changes the game configuration and plays a different preparation game with the player in the next round. We call such a game a \emph{meta-preparation game}. Similarly, one can define meta-meta preparation games, where, in each round, the referee and the player engage in a meta-preparation game. This recursive construction can be repeated indefinitely.

In Appendix~\ref{app:meta} we show that the maximum average score of a (meta)$^j$-game, which refers to a game at level $j$ of the above recursive construction, can be computed inductively, through a formula akin to eq. (\ref{induction}). Moreover, in the particular case that the preparation games that make up the (meta)$^j$-game have $\{0,1\}$ scores, one only needs to know their minimum and maximum scores to compute the (meta)$^j$-game's maximum average score.

For simple meta-games such as ``play $m$ times the $\{0,1\}$-scored preparation game $G$, count the number of wins and output $1$ ($0$) if it is greater than or equal to (smaller than) a threshold $v$'', which we denote  $G_v^{(m)}$, we find that the optimal meta-strategy for the player is to always play $G$ optimally, thus recovering 
\be
p(G,v,m)\equiv\max_{\P\in\S}G_v^{(m)}(\P)=\sum_{k=v}^m \left(\begin{array}{c}m\\k\end{array}\right)G(\P^\star)^k(1-G(\P^\star))^{m-k},
\label{binomial}
\ee
\noindent where $\P^\star=\mbox{arg}\max_{\P\in\S}G(\P)$, from~\cite{Elkouss}. $p(G,v,m)$ can be interpreted as a $p$-value for $\C$-constrained strategies, as it measures the probability of obtaining a result at least as extreme as the observed data $v$ under the hypothesis that the player's strategies are constrained to belong to $\S$.

\subsection{Devising large-round preparation games based on gradient descent} \label{sec:gradient}
\label{gradient_sec}

A more sophisticated alternative to devise many-round quantum preparation games 
exploits the principles behind Variational Quantum Algorithms \cite{VQA}. These are used to optimize the parameters of a quantum circuit by following the gradient of an operator average. Similarly, we propose a gradient-based method to identify the optimal linear witness for detecting certain quantum states. Since the resulting measurement scheme is adaptive, the techniques developed so far are crucial for studying its vulnerability with respect to an adaptive preparation attack.

Consider a set of i.i.d.\ preparation strategies ${\cal E}=\{\rho^{\otimes n}:\rho\in E\}$, and let $\{\|W(\theta)\|\leq 1:\theta\in\R^m\}\subset B(\H)$ be a parametric family of operators such that $\|\frac{\partial}{\partial\theta_x} W(\theta)\|\leq K$, for $x=1,...,m$. Given a function $f:\R^{m+1}\to \R$, we wish to devise a preparation game that, ideally, assigns to each strategy $\rho^{\otimes n}\in {\cal E}$ an average score of 
\be
f\left(\theta_\rho,\tr[W(\theta_\rho)\rho]\right), 
\label{funci}
\ee
\noindent with
\be
\theta_\rho=\mbox{argmax}_{\theta}\; \tr[W(\theta)\rho].
\label{convex_opt}
\ee
\noindent Intuitively, $W(\theta_\rho)$ represents the optimal witness to detect some property of $\rho$, and both the average value of $W(\theta_\rho)$ and the value of $\theta_\rho$ hold information regarding the use of $\rho$ as a resource.

Next, we detail a simple heuristic to devise preparation games $G$ whose average score approximately satisfies eq.\eqref{funci}. If, in addition, $f\left(\theta_\rho,\tr[W(\theta_\rho)\rho]\right)\leq \delta$ for all $\rho\in \C$, then one would expect that $G(\P)\lessapprox \delta$, for all $\C$-constrained strategies $\P\in\S$.

Fix the quantities $\epsilon>0$, $\theta_0\in\R^m$ and the probability distributions $\{p_k(x):x\in\{0,1,...,m\}\}$, for $k=1,...,n$. For $x=1,...,m$, let $\{M^x_a(\theta):a=-1,1\}$ be a POVM such that 
\be
M^x_{1}(\theta)-M^x_{-1}(\theta)=\frac{1}{K}\frac{\partial}{\partial \theta_x} W(\theta).
\label{derivative}
\ee
\noindent Similarly, let $\{M^0_{-1}(\theta), M^0_1(\theta)\}$ be a POVM such that 
\be
M^0_{1}(\theta)-M^0_{-1}(\theta) =W(\theta).
\ee 
A gradient-based preparation game then proceeds as follows. 
\begin{enumerate}
	\item
	The possible game configurations are vectors from the set $S_k = \{-(k-1),...,k-1\}^{m+1}$, for $k=1,...,n$. Given $s_k\in S_k$, we will denote by $\tilde{s}_k$ the vector that results when we erase the first entry of $s_k$.
	\item
	At round $k$, the referee samples the random variable $x\in\{0, 1,...,m\}$ from $p_k(x)$. The referee then implements the physical POVM $M^x_a(\theta_k)$, with $\theta_k=\theta_0 + \epsilon \tilde{s}_k$, obtaining the result $a_k\in\{-1,1\}$. The next game configuration is $s_{k+1} = s_k + a_k\ket{x}$.
	\item
	The final score of the game is $f\left(\theta_n,\frac{s_n^0}{\sum_{k=1}^n p_k(0)}\right)$. 
\end{enumerate}

More sophisticated variants of this game can, for instance, let $\epsilon$ depend on $k$, or take POVMs with more than two outcomes into account. It is worth remarking that, for fixed $m$, the number of possible game configurations scales with the total number of rounds $n$ as $O(n^{m+1})$.

If the player uses an i.i.d.\ strategy, then the sequence of values $(\theta_k)_k$ reflects the effect of applying stochastic gradient descent \cite{subgradient} to solve the optimization problem \eqref{convex_opt}. Hence, for the i.i.d.\ strategy $\rho^{\otimes n}$ and $n\gg 1$, one would expect the sequence of values $(\theta_k)_{k}$ to converge to $\theta_\rho$, barring local maxima. In that case, the average score of the game will be close to \eqref{funci} with high probability. For moderate values of $n$, however, it is difficult to anticipate the average game scores for strategies in ${\cal E}$ and $\S$, so that a detailed analysis with the procedure from eq. (\ref{induction}) becomes necessary (see the applications below for an example).

\section{Entanglement certification as a preparation game} \label{sec:entanglement_intro}
A paradigmatic example of a preparation game is entanglement detection. In this game, the player is an untrusted source of quantum states, while the role of the referee is played by one or more separate parties who receive the states prepared by the source. The goal of the referee is to make sure that the source has indeed the capacity to distribute entangled states. The final score of the entanglement detection preparation game is either $1$ (certified entanglement) or $0$ (no entanglement certified), that is, $g:\bar{S}\to \{0,1\}$. In this case, one can identify the final game configuration with the game score, i.e., one can take $\bar{S}=\{0,1\}$. The average game score is then equivalent to the probability that the referee certifies that the source can distribute entangled states.

Consider then a player who is limited to preparing separable states, i.e., a player for whom $\C$ corresponds to the set of fully separable states. Call the set of preparation strategies available to such a player ${\cal S}$. Ideally, we wish to implement a preparation game such that the average game score of a player using strategies from ${\cal S}$ (i.e., the probability that the referee incorrectly labels the source as entangled) is below some fixed value $e_{I}$. In hypothesis testing, this quantity is known as type-I error. At the same time, if the player follows a class $\E$ of preparation strategies (involving the preparation of entangled states), the probability that the referee incorrectly labels the source as separable is upper bounded by $e_{II}$. This latter quantity is called type-II error. In summary, we wish to identify a game $G$ such that $p(1|\P)\leq e_{I}$, for all $\P\in {\cal S}$, and $p(0|\P)\leq e_{II}$, for all $\P\in {\cal E}$.

In the following, we consider three types of referees, with access to the following sets of measurements:
\begin{enumerate}
	\item
	Global measurements: ${\cal M}_1$ denotes the set of all bipartite POVMs.
	\item
	1-way Local Pauli measurements and Classical Communication (LPCC): ${\cal M}_2$ is the set of POVMs  conducted by two parties, Alice and Bob, on individual subsystems, where Alice may perform a Pauli measurement first and then, depending on her inputs and outputs, Bob chooses a Pauli measurement as well. The final outcome is a function of both inputs and outcomes.

	\item
	Local Pauli measurements: ${\cal M}_3$ contains all POVMs where Alice and Bob perform Pauli measurements $x,y$ on their subsystems, obtaining results $a,b$, respectively. The overall output is $\gamma=f(a,b,x,y)$, where $f$ is a (non-deterministic) function. 

\end{enumerate}

\subsection{Few-round protocols for entanglement detection}

We first consider entanglement detection protocols with just a single round ($n=1$). Let $E=\{\rho_1,...,\rho_{r-1}\}$ be a set of $r-1$ bipartite entangled states. Our objective is to minimise the type-II error, given a bound $e_I$ on the acceptable type-I error. To express this optimization problem as in \eqref{optim}, we define $\S_i\equiv\{\rho_i\}$, for $i=1,...,r-1$, and $\S_r\equiv\S$, the set of separable strategies. In addition, we take $f(v)=v_1$ and choose $A,b$ so that $v_r=e_I,v_1=...v_{r-1}$. Finally, we consider complementary score functions $g,g':\bar{S}\to\{0,1\}$ and assign the scores $g_i=g$ for $i=1,...,r-1$, and $g_r=g'$. All in all, the problem to solve is
\begin{align}
&\min_{(M^{(1)}_{s|\emptyset})_s, e_{II}} e_{II}\nonumber\\
\mbox{s.t. }
&\tr(M^{(1)}_{0|\emptyset}\rho_i)\leq e_{II}, \quad i=1,...r-1,\nonumber\\
&e_I \id -M^{(1)}_{1|\emptyset}\in \C^*,\nonumber\\
&(M^{(1)}_{s|\emptyset})_s\in {\cal M}.
\label{one-shot}
\end{align}

To optimize over the dual ${\cal C^*}$ of the set of separable states, as required in \eqref{one-shot}, we invoke the Doherty-Parillo-Spedalieri (DPS) hierarchy~\cite{DPS1, DPS2}. As shown in the Appendix~\ref{app:dps}, the dual of this hierarchy approximates the set of all entanglement witnesses from the inside and converges as $n \rightarrow \infty$. In the case of two qubits the DPS hierarchy already converges at the first level. Hence, the particularly simple ansatz 
\be
e_I \id -\sum_{s\in S_2}M^{(1)}_{s|\emptyset}\langle g(s)\rangle  = V_0 + V_1^{\T_B},
\label{PPT}
\ee
where $V_0, V_1  \geq 0$ and $^{\T_B}$ is the partial transpose over the second subsystem, already leads us to derive tight bounds on the possible $e_{II}$, given $e_I$ and the class of measurements available to the referee. For larger dimensional systems, enforcing condition (\ref{PPT}) instead of the second constraint in (\ref{one-shot}) results in a sound but perhaps suboptimal protocol (namely, a protocol not necessarily minimizing $e_{II}$). Nevertheless, increasing the level of the DPS dual hierarchy generate a sequence of increasingly better (and sound) protocols whose type-II error converges to the minimum possible value asymptotically.

Eq. (\ref{one-shot}) requires us to enforce the constraint $(M^{(1)}_{s|\emptyset})_s\in \M$. For $\M=\M_1$, this amounts to demanding that the matrices $(M^{(1)}_{s|\emptyset})_s$ are positive semidefinite and add up to the identity. In that case, problem (\ref{one-shot}) can be cast as a semidefinite program (SDP) \cite{sdp}. 

For the cases $\M=\M_2,\M_3$, denote Alice and Bob's choices of Pauli measurements by $x$ and $y$, with outcomes, $a$, $b$ respectively, and call $\gamma\in\{0,1\}$ the outcome of the $1$-way LPCC measurement. Then we can express Alice and Bob's effective POVM as
\be
M^{(1)}=\sum_{a,b,x,y}P(x,y,\gamma|a,b)A_{a|x}\otimes B_{b|y},
\ee
\noindent where the distribution $P(x, y, \gamma|a,b)$ is meant to model Alice and Bob's classical processing of the outcomes they receive. For $\M=\M_2$, $P(x,y,\gamma|a,b)$ must satisfy the conditions
\footnote{Given any such $P$, Alice and Bob would conduct their joint measurement as follows: first, Alice chooses her physical setting $x$ by sampling the distribution $P(x)$. She obtains the measurement result $a$. Next, Bob chooses his setting $y$ by sampling $P(y|a,x)\equiv\frac{P(x, y|a)}{P(x)}$. He obtains the measurement result $b$. The joint measurement's effective outcome $\gamma$ is chosen by sampling over the distribution $P(\gamma|a,x,y,b)\equiv \frac{P(x,y,\gamma|a,b)}{P(x, y|a)}$.}

	\be
	\sum_{y,\gamma}P(x, y, \gamma|a,b)= P(x) \quad \text{and} \quad  \sum_{\gamma}P(x, y, \gamma|a,b)= P(x, y| a),
	\ee
\noindent whereas, for $\M=\M_3$, $P(x,y,\gamma|a,b)$ satisfies
\footnote{In this case, Alice and Bob receive measurement settings $x,y$, sampled from $P(x,y)$. Next, they obtain the measurement results $a,b$ and sample the measurement's effective outcome $\gamma$ from the distribution $P(\gamma|x,a,y,b)\equiv\frac{P(x, y, \gamma|a,b)}{P(x,y)}$.} 

	\be
	\sum_{\gamma}P(x, y, \gamma|a,b)= P(x,y).
	\ee

For $\M=\M_2,\M_3$, enforcing the constraint $(M^{(1)}_{s|\emptyset})_s\in \M$ thus requires imposing a few linear constraints on the optimization variables $P(x,y,\gamma|a,b)$. For these cases, problem (\ref{one-shot}) can therefore be cast as an SDP as well.

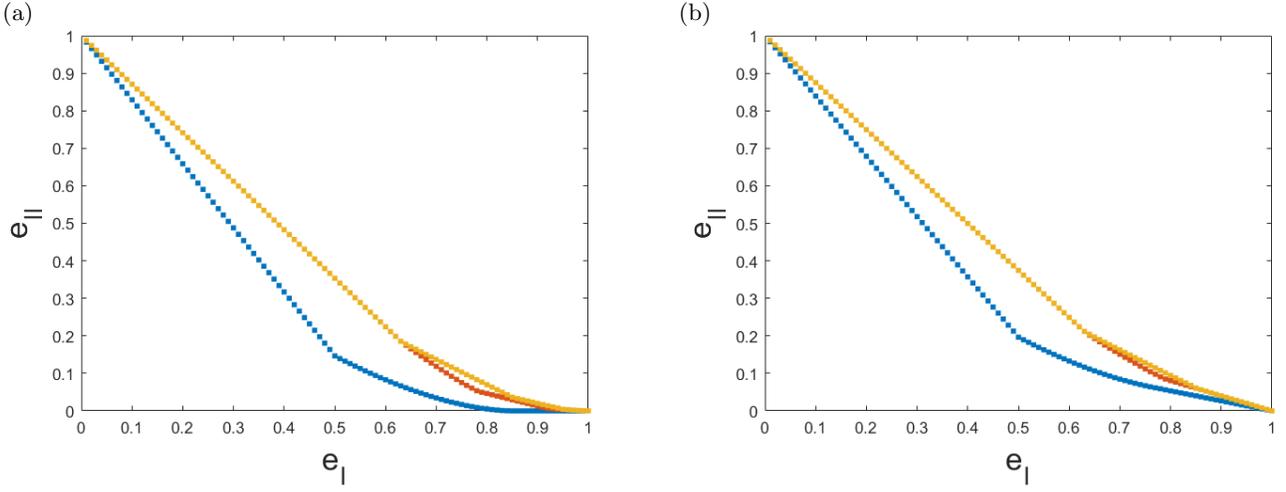
\begin{figure}
	\begin{minipage}[t]{0.485\linewidth}
		\includegraphics[width=1\textwidth]{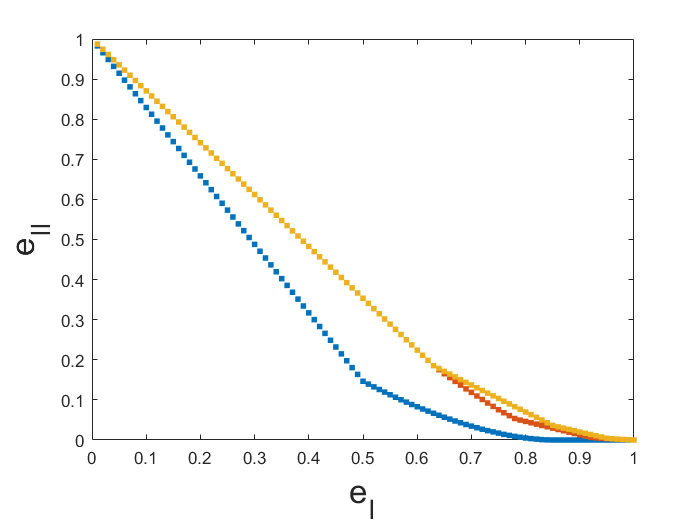}
	\end{minipage}
	\hspace{0.01\linewidth}
	\begin{minipage}[t]{0.485\linewidth}
		\includegraphics[width=1\textwidth]{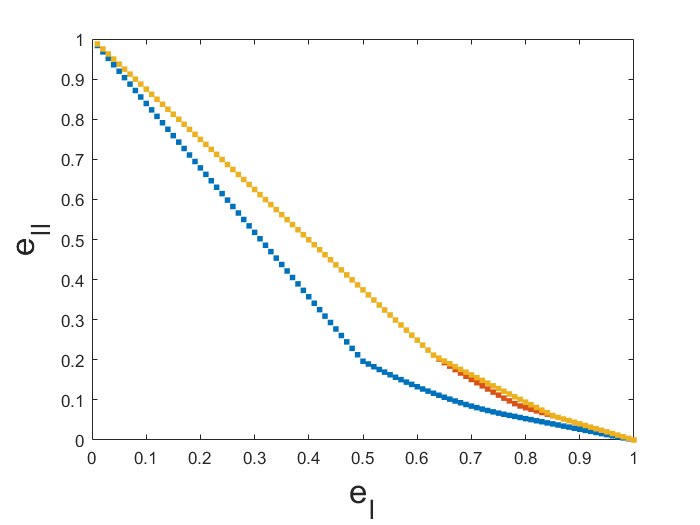}
	\end{minipage}
\begin{textblock*}{1.8cm}(0cm,-6.6cm)
	\begin{tikzpicture}
		\node at (0,0) (a) {(a)};
		\node at (9,0) (b) {(b)};
	\end{tikzpicture}
\end{textblock*}
	\caption{1-shot entanglement certification for $\ket{\phi}=\frac{1}{\sqrt{2}}(\ket{00}+\ket{1+})$. The referee has access to measurement strategies from the sets ${\cal M}_1$ (blue), ${\cal M}_2$ (red), ${\cal M}_3$ (yellow). We display the mimimal $e_{II}$ for fixed $e_I$. As each game corresponds to a hypothesis test, the most reasonable figure of merit is to quantify the type-I and type-II errors $(e_I, e_{II})$ a referee could achieve. These error pairs lie above the respective curves in the plots, any error-pair below is not possible with the resources at hand. Our optimisation also provides us with an explicit POVM, i.e., a measurement protocol, that achieves the optimal error pairs.
	(a) Entanglement detection for exact state preparation. The minimal total errors for $\ket{\phi}$ are $e_I+e_{II}=0.6464$ with ${\cal M}_1$, $e_I+e_{II}=0.8152$ with ${\cal M}_2$, and $e_I+e_{II}=0.8153$ with ${\cal M}_3$. For most randomly sampled states, these errors are much larger. We remark that there are also states, such as the singlet, where ${\cal M}_2$ and ${\cal M}_3$ lead to identical optimal errors.
	(b)	Entanglement detection for noisy state preparation. To enforce that all states close to $\rho=\proj{\psi}$ remain undetected with probability at most $e_{II}$, we need to invoke eq.~\eqref{dual_cond}, with $\C=\{\rho':\rho'\geq 0,\tr(\rho')=1,\|\rho-\rho'\|_1\leq \epsilon\}$. In Appendix~\ref{app:approx} we show how to derive the dual to this set. The plot displays the $\epsilon=0.1$ case.
 }
	\label{fig:1shot}
\end{figure}

In Figure~\ref{fig:1shot}, we compare the optimal error trade-offs for $\M=\M_1,\M_2,\M_3$ and further generalise this to scenarios, where, e.g.\ due to experimental errors, the device preparing the target state $\rho$ is actually distributing states $\epsilon$-close to $\rho$ in trace norm. The corresponding numerical optimisations, as well as any other convex optimization problem solved in this paper, were carried out using the semidefinite programming solver MOSEK \cite{mosek}, in combination with the optimization packages YALMIP \cite{yalmip} or CVX \cite{cvx}. We provide an example of a MATLAB implementation of these optimisations at~\cite{Code}.

We next consider the problem of finding the best strategy for $\M=\M_2,\M_3$ for $n$-round entanglement detection protocols. In this scenario, our general results for Maxwell demon games are not directly applicable. The reason is that, although both Alice and Bob are just allowed to conduct a finite set of physical measurements (namely, the three Pauli matrices), the set of effective local or LPCC measurements which they can enforce in each game round is not discrete. Nonetheless, a simple modification of the techniques developed for Maxwell demon games suffices to make the optimizations tractable. For this, we model Alice's and Bob's setting choices $(x_i)_i$, $(y_i)_i$ and final score $\gamma$ of the game, depending on their respective outcomes $(a_i)_i$, $(b_i)_i$ through conditional distributions
\be
P(x_1, y_1,x_2,y_2,...,x_n,y_n,\gamma|a_1,b_1,...,a_n,b_n).
\label{quid}
\ee
Depending on whether the measurements in each round are taken from $\mathcal{M}_2$ or $\mathcal{M}_3$ this distribution will obey different sets of linear constraints. For the explicit reformulation of problem (\ref{optim}) as an SDP in this setting, we refer to Appendix~\ref{app:maxwellapplications}.

\begin{figure}
	\begin{minipage}[t]{0.485\linewidth}
		\includegraphics[width=1\textwidth]{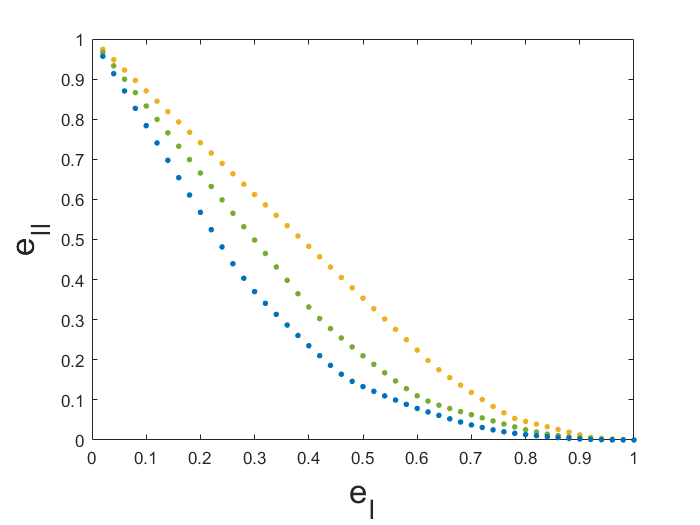} 
	\end{minipage}
	\hspace{0.01\linewidth}
	\begin{minipage}[t]{0.485\linewidth}
		\includegraphics[width=1\textwidth]{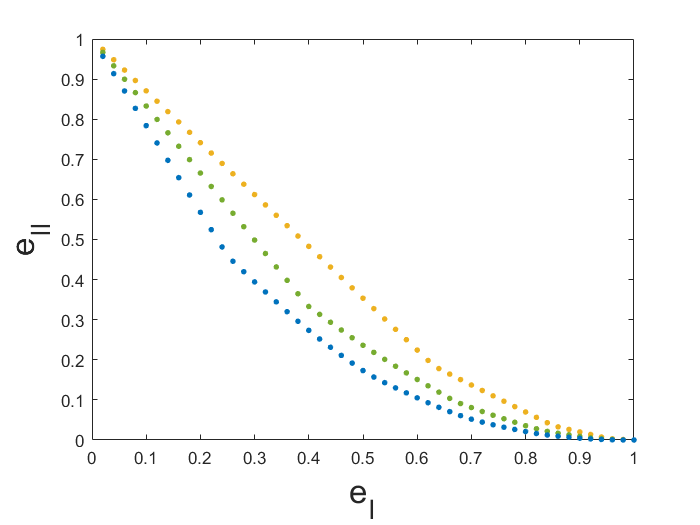} 
	\end{minipage}
	\begin{textblock*}{1.8cm}(0cm,-6.6cm)
		\begin{tikzpicture}
			\node at (0,0) (a) {(a)};
			\node at (9,0) (b) {(b)};
		\end{tikzpicture}
	\end{textblock*}
	\caption{Maxwell demon games played for various numbers of rounds. The referee has access to measurement strategies from the sets ${\cal M}_2$ (a) and ${\cal M}_3$ (b) within each round. The choice of the overall POVM implemented in each round will, in either case, depend on all inputs and outputs of previous rounds. The curves display the optimal error pairs for $n=1$ (yellow), $n=2$ (green) and $n=3$ (blue) for ${\cal E}=\{ \proj{\phi}^{\otimes n} \}$.}
	\label{fig:nshot}
\end{figure}
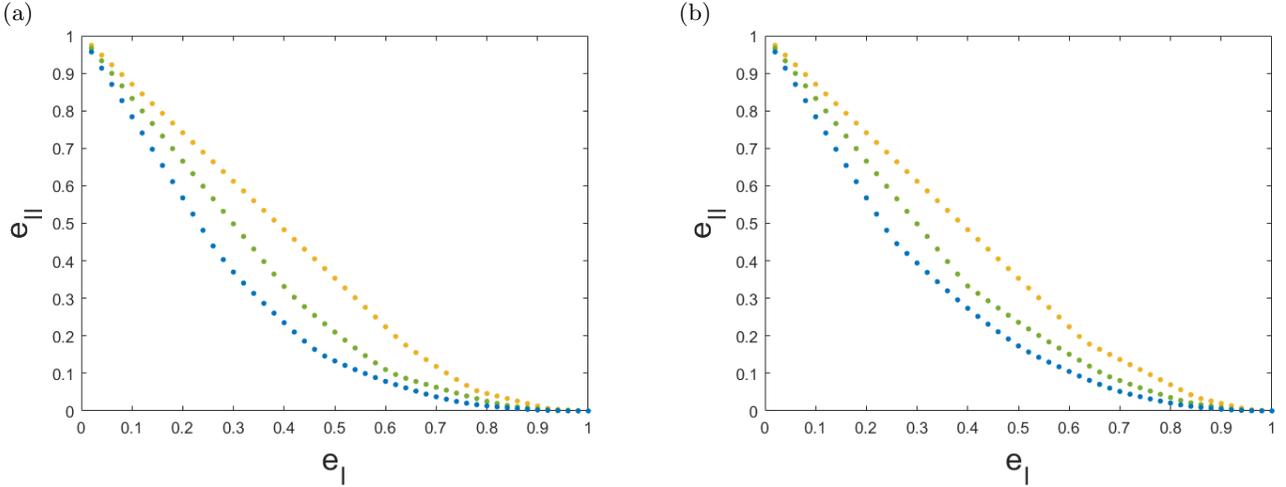

Solving this optimization problem, we find the optimal multi-round error trade-offs for two-qubit entanglement detection in scenarios where the POVMs considered within each round are either in the set ${\cal M}_2$ (LPCC) or ${\cal M}_3$ (Local Pauli measurements), see Figure~\ref{fig:nshot}. 

Now let us consider the scenario from above where within each round a measurement from class ${\cal M}_3$ is applied in more detail. Does the adaptability of the choice of POVM \emph{between} the rounds in a Maxwell demon game actually improve the error trade-offs? Specifically, we aim to compare the case where the referee has to choose a POVM from ${\cal M}_3$ for each round of the game beforehand to the case where they can choose each POVM from ${\cal M}_3$ on the fly based on their previous inputs and outputs. The answer to this question is intuitively clear when we consider a set $E$ of more than one state, since then we can conceive a strategy where in the first round we perform a measurement that allows us to get an idea which of the states in $E$ we are likely dealing with, while in the second round we can then use the optimal witness for that state. However, more surprisingly, we find that this can also make a difference for a single state $E=\{\proj{\psi}\}$. For instance, for the state $\ket{\psi}=\frac{1}{\sqrt{2}}(\ket{0+}+ \ket{1{-i}})$ with $\ket{{-i}}=\frac{1}{\sqrt{2}}(\ket{0}-i\ket{1})$, we find that, in two-round games, the minimum value of $e_I+e_{II}$ equals $0.7979$ with adaptation between rounds and $0.8006$ without adaptation (see~\cite{mateus} for a statistical interpretation of the quantity $e_I+e_{II}$).

This result may strike the reader as surprising: on first impulse, one would imagine that the best protocol to detect the entanglement of two preparations of a known quantum state $\rho$ entails testing the same entanglement witness twice. A possible explanation for this counter-intuitive phenomenon is that preparations in $\E$ and $\S$ are somehow correlated: either both preparations correspond to $\rho$ or both preparations correspond to a separable state. From this point of view, it is not far-fetched that an adaptive measurement strategy can exploit such correlations.

Our framework also naturally allows for the optimisation over protocols with  $e_{II}=0$ and where the corresponding $e_I$ error is being minimised, thus generalising previous work on detecting entanglement in few experimental rounds~\cite{Bori1, Bori2}. Using the dual of the DPS hierarchy for full separability~\cite{DPS3}, we can furthermore derive upper bounds on the errors for states shared between more than two parties. 
Similarly, a hierarchy for detecting high-dimensional entangled states~\cite{beyondfidelities} allows us to derive protocols for the detection of high-dimensional entangled states using quantum preparation games in~\cite{ExperimentalPaper}.

Due to the exponential growth of the configuration space, optimisations over Maxwell demon adaptive measurement schemes are hard to conduct even for relatively low values of $n$. Devising entanglement detection protocols for $n\gg 1$ requires completely different techniques.

\subsection{Many-round protocols for entanglement detection}

\begin{figure}
	\begin{minipage}[t]{0.6\linewidth}
		\includegraphics[width=1\textwidth]{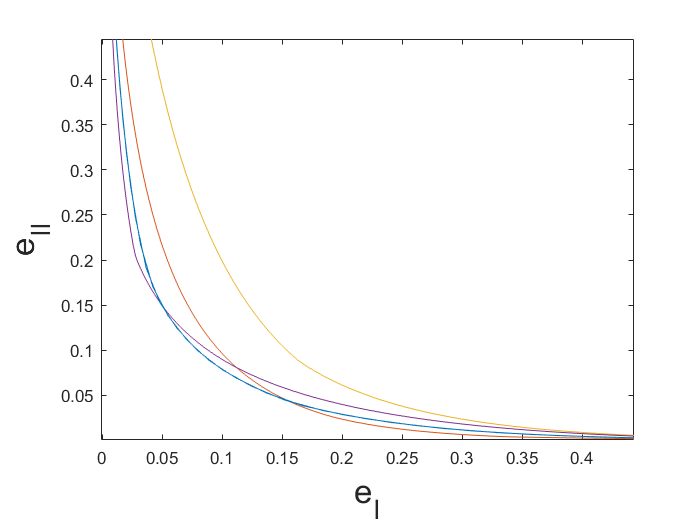}
	\end{minipage}
	\caption{Comparison of independent repetitions of $1$-shot and $3$-round games for $E=\{ \proj{\phi} \}$. 
		The games $G^{(30)}_{22}$ (yellow), $G^{(30)}_{25}$ (red) and $G^{(30)}_{28}$ (purple) are obtained through $30$ independent repetitions of optimal one-shot games $G$ restricted to measurements in $\mathcal{M}_2$. These are compared to the optimal $3$-round adaptive protocols $G'$  with measurements $\mathcal{M}_2$ performed in each of the three rounds, independently repeated $10$ times as $G'^{(10)}_{8}$ (blue). The $1$ and $3$-shot games $G$ and $G'$ are also displayed in Figure~\ref{fig:nshot}. We observe that the repetition of the adaptive protocol outperforms the others in the regime of low $e_I+e_{II}$. }
	\label{fig:binomial}
\end{figure}

In order to devise many-round preparation games, an alternative to carrying out full optimizations is to rely on game composition.\footnote{Recall that the heuristic presented in Box~2 is another viable option for moderate round numbers. We illustrate this in Appendix~\ref{app:implementationfincorr}, where we use it to devise $20$-round protocols for entanglement detection.} In this regard, in Figure~\ref{fig:binomial} we compare $10$ independent  repetitions of a $3$-round adaptive strategy to $30$ independent repetitions of a 1-shot protocol, based on~\eqref{binomial}. 
\label{sec:boundedinomialfigures}
This way of composing preparation games can easily be performed with more repetitions. Indeed, for $m=1000$ repetitions we find preparation games with errors at the order of $\approx 10^{-14}$. 
In the asymptotic regime, the binomial distribution of the number of $1$-outcomes for a player restricted to separable strategies (see eq.~\eqref{binomial}) can be approximated by a normal distribution. For $e_I < \mu\equiv \frac{v}{m}$, this leads to a scaling as $e_{I}(m) \approx e^{-\frac{m(\mu-e_I)^2}{2 e_I (1-e_I)}}$ (similarly for $(1-e_{II})$ when $\mu < 1-e_{II}$, and where the player is preparing states from $\E$).

Finally, we apply gradient descent as a guiding principle to devise many-round protocols for entanglement quantification. 
For experimental convenience, the preparation game we develop is implementable with 1-way LOCC measurements. 

We wish our protocol to be sound for i.i.d.\ strategies in ${\cal E}=\{\rho^{\otimes n}:\rho\in E\}$, with $E$ being the set of all states
\be
\ket{\psi_\theta}= \cos(\theta)\ket{00} + \sin(\theta)\ket{11},
\ee
\noindent for $\theta\in (0,\pi/2)$. For such states, the protocol should output a reasonably good estimate of $\ket{\psi_\theta}$'s entanglement entropy, $S(\ket{\psi_\theta})=h(\cos^2(\theta))$, with $h(x)=-x\log(x) -(1-x)\log(1-x)$ the binary entropy.  \del{Most} Importantly, if the player is limited to preparing separable states, the average score of the game should be low. 

Following eq. (\ref{funci}), we introduce 
\be
W(\theta)=\frac{1}{2}\left[Z \otimes Z + \proj{+}\otimes \left( \sin(2\theta)X+\cos(2\theta)Z\right) + \proj{-}\otimes \left( - \sin(2\theta)X + \cos(2\theta)Z\right)\right].
\ee
\noindent This operator satisfies $\|W(\theta)\|\leq 1$ and $\ket{\psi_\theta}$ is the only eigenvector of $W(\theta)$ with eigenvalue $1$. $W(\theta)$ can be estimated via 1-way LOCC with the POVM $M^0_{-1}(\theta)= \frac{\id - W(\theta)}{2}$, $M^0_{1}(\theta)= \frac{\id + W(\theta)}{2}$. Furthermore, consider
\be
\frac{\partial}{\partial \theta}W =  \phantom{\frac{'}{'}} \! \! \! \proj{+}\otimes \left(\cos(2\theta)X-\sin(2\theta)Z\right) - \proj{-}\otimes \left(\cos(2\theta)X + \sin(2\theta)Z\right) .
\ee
This dichotomic observable can be estimated via eq.\eqref{derivative} with the 1-way LOCC POVM defined by 
\begin{align}
&M^1_{-1}(\theta) = \proj{+}\otimes \frac{1}{2}\left(\id - \cos(2\theta)X+\sin(2\theta)Z\right) + \proj{-}\otimes \frac{1}{2}\left(\id + \cos(2\theta)X + \sin(2\theta)Z\right) , \nonumber\\
&M^1_{1}(\theta)=\id-M^1_{-1}(\theta),
\end{align}
which satisfies $M^1_{1}-M^1_{-1}= \frac{\partial}{\partial \theta}W$.

\begin{figure}
	\begin{minipage}[t]{0.44\linewidth}
		\includegraphics[width=1\textwidth]{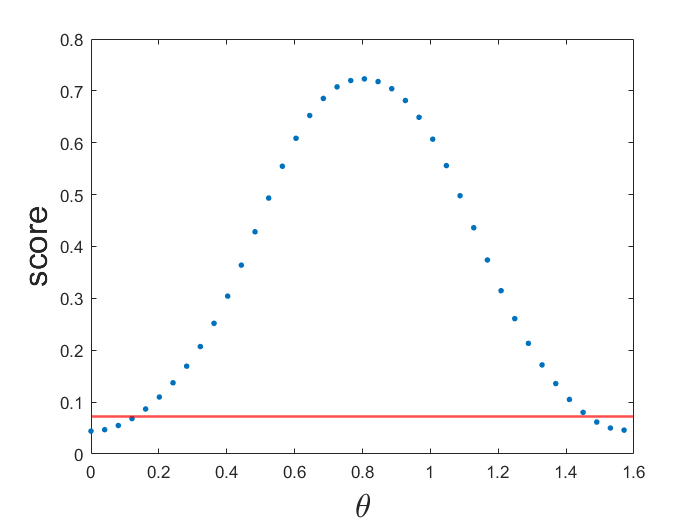}
	\end{minipage}
	\hspace{0.01\linewidth}
	\begin{minipage}[t]{0.44\linewidth}
		\includegraphics[width=1\textwidth]{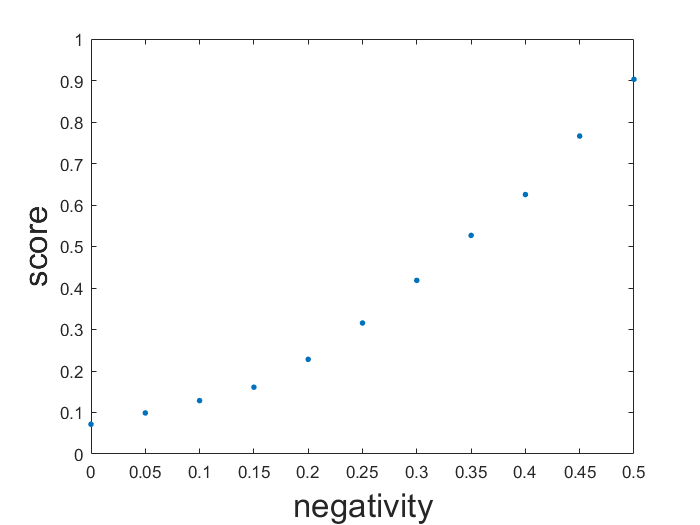}
	\end{minipage}
\begin{textblock*}{1.8cm}(0.5cm,-6.2cm)
	\begin{tikzpicture}
		\node at (0.5,0) (a) {(a)};
		\node at (9,0) (b) {(b)};
	\end{tikzpicture}
\end{textblock*}
	\caption{Gradient descent based preparation game with parameters $\epsilon=0.1$, $\lambda=0.1$ and $\theta_0=0$. The probability of measuring $\left\{M^0_1, M^0_{-1}\right\}$ in round $k$ is chosen according to $p_k(0)=\frac{1}{1+e^{-(2k-n)}}$. This captures the intuition that in the first few rounds it is more important to adjust the angle, while in later rounds the witness should be measured more often. (a) The score assigned to i.i.d.\ preparation strategies as a function of the parameter $\theta$ of $\ket{\psi_\theta}$ for $n=41$ rounds for ${\cal E}$ (blue) compared to the optimal separable value (red). As expected, the average game scores of the i.i.d.\ strategies $\{\proj{\psi_{\theta}}^{\otimes n}:\theta\}$ mimic the shape of the curve $h(\cos(\theta)^2)$ and the scores obtainable with the set of separable strategies $\S$ perform significantly worse compared to the states from ${\cal E}$ with angles close to $\theta= \frac{\pi}{4}$. (b) The optimal scores achievable by players capable of preparing bipartite quantum states of bounded negativity \cite{negativity}, obtained through application of eq.~\eqref{induction}. We observe that the average score of the game constitutes a good estimator for entanglement negativity.}
	\label{fig:grad}
\end{figure}
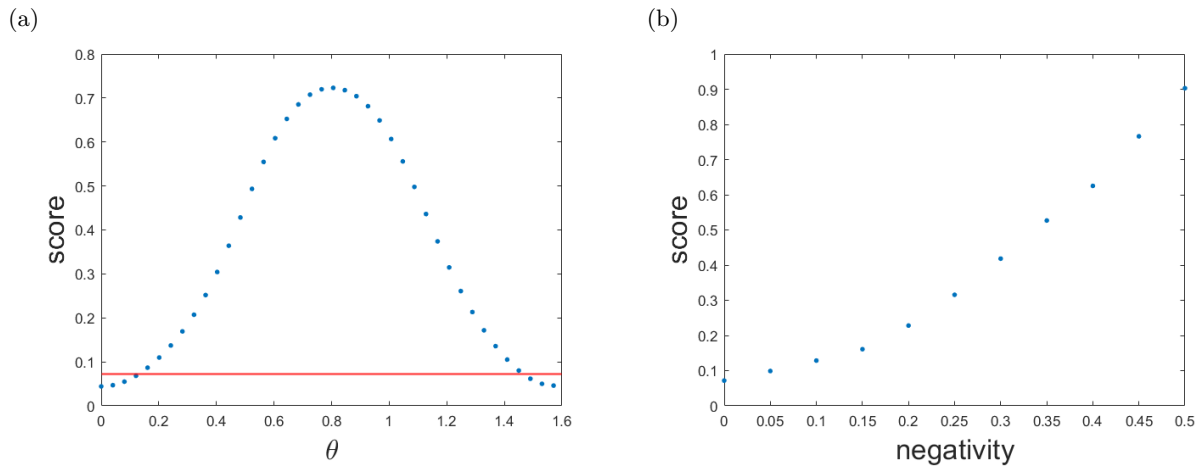

Let us further take $f(\theta, v)= h\left(\cos^2(\theta)\right)\Theta(v-(1-\lambda + \lambda\delta(\theta)))$ with $0 \leq \lambda \leq 1$ and $\delta(\theta)=\max_{\rho \in \C} \tr[W(\theta)\rho]$. This captures the following intuition: if the estimate $v$ of $\tr[W(\theta_n)\rho]$ is below a convex combination of the maximum value achievable (namely, $\bra{\psi_{\theta}}W(\theta_n = \theta)\ket{\psi_{\theta}}=1$) and the maximum value $\delta(\theta_n)$ achievable by separable states, then the state shall be regarded as separable and thus the game score is set to zero. In Figure \ref{fig:grad}, we illustrate how this game performs. 

\section{Conclusion}

We have introduced quantum preparation games as a convenient framework to analyze the certification and quantification of resources. We derived general methods to compute the (maximum) average score of arbitrary preparation games under different restrictions on the preparation devices: this allowed us to prove the soundness or security of general certification protocols. Regarding the generation of such protocols, we explained how to conduct exact (approximate) optimizations over preparation games with a low (moderate) number of rounds. In addition, we introduced two methods to devise large-round preparation games, via game composition and through gradient descent methods. These general results were applied to devise novel protocols for entanglement detection and quantification. To our knowledge, these are the first non-trivial adaptive protocols ever proposed for this task. In addition, we discovered that, against the common practice in entanglement detection, entanglement certification protocols for a known quantum state can often be improved using adaptive measurement strategies.

Even though we illustrated our general findings on quantum preparation games with examples from entanglement theory, where the need for efficient protocols is imminent, we have no doubt that our results will find application in other resource theories. With the current push towards building a quantum computer, a second use of our results that should be particularly emphasized is the certification of magic states. More  generally, developing applications of our work to various resource theories, including for instance the quantification of non-locality, is an interesting direction for future work.

Another compelling line of research consists in studying the average performance of preparation games where Assumption 1 does not hold. In those games, a player can exploit the action of the referee's measurement device to generate states outside the class allowed by their preparation device. Such games naturally arise when the player is limited to preparing resource-free states for some resource theory, but the referee is allowed to conduct resourceful measurements. An obvious motivating example of these games is the detection of magic states via general POVMs.

Finally, it would be interesting to explore an extension of preparation games where the referee is allowed to make the received states interact with a quantum system of  fixed dimension in each round. This scenario perfectly models the computational power of a Noisy Intermediate-Scale Quantum (NISQ) device. In view of recent achievements in experimental quantum computing, this class of games is expected to become more and more popular in quantum information theory.

\acknowledgements
This work was supported by the Austrian Science fund (FWF) stand-alone project P~30947. This is a preprint of an article published in Nature Communications. The final authenticated version is available online at:\newline
https://doi.org/10.1038/s41467-021-24658-9.

\appendix

\section{The maximum average score of finitely correlated strategies}
\label{app:fincorr}
\label{app:fincorrconstr}

Here we explain how to compute the maximum average score achievable in a preparation game by a player conducting a finitely correlated strategy (see Figure~\ref{fig:player}), under the assumption that the quantum operation effected by the preparation device is known, but not the initial state of the environment. 

In such preparations, the player's device interacts with an environment $A$. More specifically, in each round, the referee receives a state
$$
\tr_{A}\left[ \sum_i K_i \rho K_i^\dagger\right],$$
where $\rho$ is the current state of the environment and $K_i:\H_A\to \H_A\otimes\H$ are the Kraus operators, which evolve the environment and prepare the state that the referee receives. Since the same environment is interacting with each prepared state, the states that the referee receives in different rounds are generally correlated.

Suppose that the referee concludes the first round of their adaptive strategy in the game configuration $s$. The (non-normalized) state of the environment will then be
\be
\sum_{i_1,j_1,l_1}\bra{l_1}M^{(1)}_{s|\emptyset}\ket{j_1}\tilde{K}_{i_1,j_1}\rho (\tilde{K}_{i_1,l_1})^\dagger,
\ee
\noindent where $\tilde{K}_{ij} = \left(\id_A\otimes\bra{j}\right)K_i$. Iterating, we find that, if the referee observes the sequence of game configurations $\emptyset,s_2,...,s_{n},\bar{s}$, then the final state of the environment will be
\be
\sum_{\vec{i},\vec{j},\vec{l}}\bra{l_1}M^{(1)}_{s_2|\emptyset}\ket{j_1}...\bra{l_n}M^{(n)}_{\bar{s}|s_{n}}\ket{j_n}\tilde{K}_{i_n,j_n}...\tilde{K}_{i_1,j_1}\rho(\tilde{K}_{i_1,l_1})^\dagger...(\tilde{K}_{i_n,l_n})^\dagger.
\ee
The probability to obtain such a sequence of configurations is given by the trace of the above operator. The average score of the game is thus $\tr[\rho\Omega]$, where the operator $\Omega$ is defined by:
\be
\Omega = \sum_{s_2,...,s_{n},\bar{s}} \ \sum_{\vec{i},\vec{j},\vec{l}}\bra{l_1}M^{(1)}_{s_2|\emptyset}\ket{j_1}...\bra{l_n}M^{(n)}_{\bar{s}|s_{n}}\ket{j_n}(\tilde{K}_{i_1,l_1})^\dagger...(\tilde{K}_{i_n,l_n})^\dagger\tilde{K}_{i_n,j_n}...\tilde{K}_{i_1,j_1}\langle g(\bar{s})\rangle.
\label{omega_def}
\ee
Note that $\Omega$ can be expressed as the composition of a sequence of linear transformations. More concretely, consider the following recursive definition
\begin{align}
&\Omega^{(n)}_{s} =\sum_{\bar{s}\in \bar{S}}\sum_{i,j, l}(\tilde{K}_{i,j})^\dagger\tilde{K}_{i,l}\bra{l}M^{(n)}_{\bar{s}|s}\ket{j}\langle g(\bar{s})\rangle,\nonumber\\
&\Omega^{(k)}_s =\sum_{i,j,l}\sum_{s'\in S_{k+1}}(\tilde{K}_{i,j})^\dagger \Omega^{(k+1)}_{s'}\tilde{K}_{i,l}\bra{l}M^{(k)}_{s'|s}\ket{j}.
\label{recursive_omega}
\end{align}
\noindent Then it can be verified that $\Omega = \Omega^{(1)}_{\emptyset}$. Calling $D$ the Hilbert space dimension of the environment, the average score of the considered preparation game can thus be computed with $O\left(D^2\sum_{k}|S_k||S_{k+1}|\right)$ operations.

In realistic experimental situations, the player will not know the original quantum state $\rho_A$ of the environment. In that case, we may be interested in computing the maximum average score achievable over all allowed environment states. Let us assume that $\rho_A\in {\cal A}$, for some convex set ${\cal A}$. Then, the maximum average score is
\be
\max_{\rho_A\in {\cal A}}\tr[\rho_A\Omega].
\label{max_score_correlated}
\ee
\noindent In case the environment is fully unconstrained, this quantity equals the maximum eigenvalue of $\Omega$.

This condition can be seen to be equivalent to
\be
v\id-\Omega\in {\cal A}^*,
\label{envi_cond}
\ee
\noindent where ${\cal A}^*$ denotes the dual of ${\cal A}$, i.e., ${\cal A}^*=\{X:\tr(X\rho)\geq 0,\forall\rho\in {\cal A} \}$. In the particular case where the initial state of the environment is unconstrained, the condition turns into
\be
v\id-\Omega\geq 0.
\label{envi_cond_pos}
\ee
\noindent Since $\Omega$ is a linear function of the optimization variables, condition \eqref{envi_cond} -- or \eqref{envi_cond_pos} -- is a convex constraint and thus we can handle it within the framework of convex optimization theory.

\section{Enforcing $\mathcal{C}$-constrained preparation strategies in Maxwell-demon games} \label{app:maxwelldemon}
In the following we show how to turn~\eqref{reduction} into a set of linear constraints on $\{P(y_0,...,y_{n}|a_0,...,a_n):a_0,...,a_n\}$. We then show how to formulate the constraints \eqref{upper_score}, when $\S$ is a set of $\C$-constrained strategies in terms of the variables $\{P(y_0,...,y_{n}|a_0,...,a_n):a_0,...,a_n\}$. This allows us to treat the optimization of multi-round Maxwell demon games with convex optimization techniques.
	
Let us first show that \eqref{reduction} and \eqref{NSP} are equivalent. That any distribution of the form \eqref{reduction} satisfies \eqref{NSP} can be checked with a straightforward calculation. 
Conversely, for any set of distributions $\{P(y_0,...,y_{n}|a_0,...,a_n):a_0,...,a_n\}$ satisfying \eqref{NSP}, there exist distributions $P_k(x_k|s_k)$, $P_{n+1}(s_{n+1}|s_{n}, a_n, x_n)$ such that \eqref{reduction} holds~\cite{costantino}. Indeed, one can derive the latter from  $\{P(x_1,...,x_k|a_0, a_1,...,a_{k-1})\}_k$ via the relations
\begin{align}
&P_k(x_k|s_k)=\frac{P(x_1,...,x_k|a_0,...,a_{k-1})}{P(x_1,...,x_{k-1}|a_0,...,a_{k-2})}\nonumber\\
&P(\score|s_{n+1})=\frac{P(x_1,...,x_n, \score|a_0, a_1,...,a_n)}{P(x_1,...,x_n|a_0, a_1,...,a_{n-1})}.
\end{align}
\noindent For fixed measurements $\{N^{(k)}_{a|x}:a,x\}$, optimizations over Maxwell demon games thus reduce to optimizations over non-negative variables $P(x_1,...,x_n, \score|a_0, a_1,...,a_n)$ satisfying eq.\eqref{NSP}, positivity and normalization
\be
\sum_{y_0,...,y_n}P(y_0,...,y_{n}|a_0,...,a_n)=1 \quad \forall a_0,...,a_n.
\label{normalization}
\ee

\bigskip

We next show how to enforce the constraint \eqref{upper_score} when $\S$ corresponds to the set of $\C$-constrained preparation strategies, for some set of states $\C$. Similarly to \eqref{induction}, we can enforce this constraint inductively. For $k=1,...,n$, let $\nu^{(k)}_{s_k}$, $\xi_{{s}_{n+1}}$ be optimization variables, satisfying the linear constraints
	\begin{align}
	&\xi_{s_{n+1}}=\sum_{\score\in \G}\score P(x_1,...,x_n,\score|a_0,a_1,...,a_n),\label{initial1}\\
	&\nu^{(n)}_{s_n}\id-\sum_{a_{n},x_{n}}\xi_{s_{n+1}}N^{(n)}_{a_{n}|x_{n}}\in \C^*,
	\label{initial2}
	\end{align}
	\noindent and 
	\be
	\nu^{(k)}_{s_k}\id-\sum_{a_{k},x_{k}}\nu^{(k+1)}_{s_{k+1}}N^{(k+1)}_{a_{k+1}|x_{k+1}}\in \C^*.
	\label{EW}
	\ee
	We claim that $\nu^{(1)}_{\emptyset}$ is an upper bound on the maximum average score achievable by a player restricted to prepare states in $\C$. Indeed, let $\rho^{(k)}_{s_k}\in \C$ be the player's preparation at stage $k$ conditioned on the game configuration $s_k$. Multiply eq.\eqref{initial2} by $\rho^{(n)}_{s_n}$ and take the trace. Then, since eq.\eqref{initial2} belongs to the dual set of $\C$, we have that
	\be
	\nu^{(n)}_{s_n}\geq\sum_{a_{n},x_{n}}\sum_{\score\in \G}\score P(x_1,...,x_n,\score|a_0,a_1,...,a_n) \tr\left[N^{(n)}_{a_{n}|x_{n}}\rho^{(n)}_{s_n}\right].
	\ee
	Next, we multiply both sides of the above equation by $\tr(N^{(n-1)}_{a_{n-1}|x_{n-1}}\rho^{(n-1)}_{s_{n-1}})$ and sum over the variables $a_{n-1},x_{n-1}$. By eq.\eqref{EW}, the result will be upper bounded by $\nu^{(n-1)}_{s_{n-1}}$. Iterating this procedure, we arrive at
	\be
	\nu^{(1)}_{\emptyset}\geq \sum_{a_1,...,a_n,x_1,...,x_n}\sum_{\score\in\G}\score P(x_1,...,x_{n},\score|a_0,a_1,...,a_{n})\prod_{k=1}^n\tr[N^{(k)}_{a_{k}|x_{k}}\rho^{(k)}_{s_k}].
	\ee
	\noindent The right-hand side is the average score of the game.
	Call $\omega^{(k)}_{s_k}\in\C^*$ the operator expressions appearing in eqs.~\eqref{initial2}, \eqref{EW}. Note that, if there exist states $\rho^{(k)}_{s_k}\in\C$ such that $\tr(\omega^{(k)}_{s_k}\rho^{(k)}_{s_k})=0$, i.e., if all the dual elements are tight, then the preparation strategy defined through the states $\{\rho^{(k)}_{s_k}\}$ achieves the average score $\nu^{(1)}_\emptyset$.

In sum, optimizations of the sort \eqref{optim} over the set of all Maxwell demon games require optimizing over $P$ under non-negativity and the linear constraints \eqref{NSP}, \eqref{normalization}. Constraints of the form \eqref{upper_score} for $\S=\{\P\}$ translate as extra linear constraints on $P$ and the upper bound variable $v$. When $\S$ corresponds to a finitely correlated strategy with unknown environment state, we can formulate condition \eqref{upper_score} as the convex constraint \eqref{envi_cond}. Finally, when $\S$ corresponds to a set of $\C$-constrained strategies, condition \eqref{upper_score} is equivalent to enforcing constraints \eqref{initial1}, \eqref{initial2} and \eqref{EW} on $P$ and the slack variables $\nu, \xi_{s_{n+1}}$, with $v\equiv \nu^{(1)}_\emptyset$.

\section{Computing the average score of a meta-preparation game}
\label{app:meta}

Our starting point is an $n$-round meta-game with configuration spaces $S=(S_1,S_2,...,S_{n+1})$, with $S_1=\{\emptyset\}$. In each round $k$, the referee runs a preparation game $G_k(s_k)$. Depending on the outcome $o_k\in O_k$ of the preparation game, the referee samples the new configuration $s_{k+1}$ from the distribution $c_k(s_{k+1}|o_k, s_k)$. The final score of the meta-game is decided via the non-deterministic function $\gamma:S_{n+1}\to\R$.

To find the optimal score achievable by a player using $\C$-constrained strategies, we proceed as we did for preparation games. Namely, define $\nu_s^{(k)}$ as the maximum average score achievable by a $\C$-constrained player, conditioned on the game being in configuration $s\in S_k$ at round $k$. Then we see that

\begin{align}
&\nu_s^{(n)}=\max_{\P\in\S}\sum_{o,s'}p(o|\P,G_n(s))c_n(s'|s,o)\langle \gamma(s')\rangle,\nonumber\\
&\nu^{(k)}_s=\max_{\P\in\S}\sum_{o,s'}p(o|\P,G_k(s))c_k(s'|s,o)\nu_{s'}^{(k+1)}.
\label{induction2}
\end{align}

Note that the first optimization above consists in finding the maximum average score of the preparation game $G_n(s)$ with score function $g(o)=\sum_{s'}c_n(s'|s,o)\gamma(s')$. Similarly, the second optimization corresponds to computing the maximum score of a preparation game with score function $g(o)=\sum_{s'}c_k(s'|s,o)\nu_{s'}^{(k+1)}$. Applying formula (\ref{induction}) iteratively, we can thus compute the average score of the meta-preparation game through $O(nn')$ operations, where $n'$ denotes the maximum number of rounds of the considered preparation games.

Notice as well that formula (\ref{induction2}) also applies to compute the score of a meta$^j$-preparation game, if we understand $\{G_k(s):s\in S_k,k\}$ as meta$^{j-1}$-preparation games.

Now, consider a scenario where the games (or meta$^{j-1}$-games) have just two possible final configurations, i.e., $o\in\{0,1\}$. In that case,

\begin{align}
&\nu_s^{(n)}=\Gamma^{(n)}_{s,1} + \max_{\P\in\S}p(0|\P,G_n(s))(\Gamma^{(n)}_{s,0}-\Gamma^{(n)}_{s,1}),\nonumber\\
&\nu^{(k)}_s=\Gamma^{(k)}_{s,1}+\max_{\P\in\S}p(0|\P,G_k(s))(\Gamma^{(k)}_{s,0}-\Gamma^{(k)}_{s,1}),
\end{align}
\noindent where
\be
\Gamma^{(n)}_{s,o}\equiv \sum_{s'}c_n(s'|s,o)\langle g(s')\rangle, \quad \Gamma^{(k)}_{s,o}\equiv \sum_{s'}c_k(s'|s,o)\nu^{(k+1)}_{s'},
\ee
\noindent for $o=0,1$.

Call $p_{\max}(G)$ ($p_{\min}(G)$) the solution of the problem $\max_{\P\in\S} p(0|G,\P)$ ($\min_{\P\in\S} p(0|G,\P)$). Then we have that 

\begin{align}
\nu_s^{(n)}&=\begin{cases}\Gamma^{(n)}_{s,1} + p_{\max}(G_n(s))(\Gamma^{(n)}_{s,0}-\Gamma^{(n)}_{s,1}), &\mbox{ for } \Gamma_{s,0}^{(n)} > \Gamma_{s,1}^{(n)}, \nonumber\\
\Gamma^{(n)}_{s,1} + p_{\min}(G_n(s))(\Gamma^{(n)}_{s,0}-\Gamma^{(n)}_{s,1}), &\mbox{ otherwise},\nonumber\\
\end{cases} \\
\nu^{(k)}_s &= \begin{cases} \Gamma^{(k)}_{s,1} + p_{\max}(G_k(s))(\Gamma^{(k)}_{s,0}-\Gamma^{(k)}_{s,1}), &\mbox{ for } \Gamma_{s,0}^{(k)} > \Gamma_{s,1}^{(k)},\nonumber\\
\Gamma^{(k)}_{s,1} + p_{\min}(G_k(s))(\Gamma^{(k)}_{s,0}-\Gamma^{(k)}_{s,1}), &\mbox{
	otherwise}.
\end{cases}
\end{align}

If the same set $J$ of meta$^{j-1}$-games are re-used at each round of the considered meta$^{j}$-game, this formula saves us the trouble of optimizing over meta$^{j-1}$-games for every round $k$ and every $s\in S_k$. The complexity of computing the maximum average score is, in this case, of order $O(n) + 2|J|c$, where $c$ is the computational cost of optimizing over a meta$^{j-1}$-game.

Think of a meta$^{j}$-game where a given meta$^{j-1}$-game $G$ (with scores $o\in\{0,1\}$) is played $m$ times, $s_k\in\{0,...,k-1\}$ corresponds to the number of $1$'s obtained, and success is declared if $s_{m+1}>v$, for some $v\in\{0,...,m\}$. Then, we have that 

\be
\Gamma_{s,o}^{(n)}=\Theta(s+o-v),\Gamma_{s,o}^{(k)}=\nu_{s+o}^{(k+1)},
\ee
\noindent where $\Theta(x)=0$, for $x<0$ or $1$ otherwise. It is thus clear that $\Gamma_{s,0}^{(k)}\leq \Gamma_{s,1}^{(k)}$ for all $k$, and so the best strategy consists in always playing to maximize $p(1|G,\P)$ in each round. In turn, this implies the binomial formula (\ref{binomial}) derived in \cite{Elkouss}.

Furthermore, as shown in \cite{mateus}, if the player uses a strategy ${\cal Q}\not\in \S$ to play $G$, with $G({\cal Q})\geq G(\P^\star)$, then the average value of $p(G,v,m)$ can be seen to satisfy
\be 
\sum_{v=0}^m p^{(m)}(v|{\cal Q})p(G,v,m)\leq \left[1-\left(G({\cal Q})-G(\P^\star)\right)^2\right]^m,
\label{eq:pvalue}
\ee
\noindent where $p^{(m)}(v|{\cal Q})$ denotes the probability of winning $v$ times with strategy ${\cal Q}$. This relation has important applications for hypothesis testing: if, by following the strategy ${\cal Q}$, we wish to falsify the hypothesis that the player is using a strategy in $\P$, all we need to do is play a preparation game for which $G({\cal Q})-G(\P^\star)$ is large enough multiple times.

\section{Optimizing over the set of separable states and its dual} \label{sec:duals}
\label{app:dps_higher}
\label{app:dps}
In the main text, we frequently encountered convex constraints of the form
\be
v\id -W\in \C^*,
\label{dualita}
\ee
\noindent where $W$ is an operator and $\C$ is a convex set of quantum states. Furthermore, we had to conduct several optimizations of the form
\be
f^\star=\max_{\rho\in \C}\tr[W\rho].
\label{maxi}
\ee
In the following, we will explain how to tackle these problems when $\C$ corresponds to the set $\texttt{SEP}$ of separable quantum states on some bipartite Hilbert space $\H_A \otimes \H_B$. 

In this regard, the Doherty-Parrilo-Spedalieri (DPS) hierarchy~\cite{DPS1, DPS2} provides us with a converging sequence of semi-definite programming outer approximations to $\texttt{SEP}$. Consider the set $E_k$ of $k+1$-partite quantum states defined by
\be
E_k=\{\rho_{AB_1\ldots B_k} : \Pi_k \rho_{AB_1\ldots B_k} \Pi_k = \rho_{AB_1\ldots B_k}, \  \rho_{AB_1\ldots B_k}^{\T_S} \geq 0 \ \forall \ S \in {\cal N} \text{ and } \tr[\rho_{AB_1\ldots B_k}]=1 \},
\label{def_E}
\ee
where $\Pi_k$ is the projector onto the symmetric subspace of $\H_{B_1} \otimes \cdots \otimes \H_{B_k}$; ${\cal N}$ is the power set of $\{B_1, \ldots B_k \}$; and $^{\T_S}$ denotes the partial transpose over the subsystems $S$.

We say that the quantum state $\rho_{AB} $ admits a \emph{Bose-symmetric PPT extension to $k$ parts on system $B$} iff there exists $\rho_{AB_1\ldots B_k}\in E_k$ such that $\rho_{AB}=\tr_{B_2,...,B_k}(\rho_{AB_1\ldots B_k})$. Call $\texttt{SEP}^k$ the set of all such bipartite states. Note that the condition $\rho_{AB}\in\texttt{SEP}^k$ can be cast as a semidefinite programming constraint.

As shown in \cite{DPS1, DPS2}, $\texttt{SEP}^1\supset\texttt{SEP}^2\supset...\supset \texttt{SEP}$ and $\lim_{k\to\infty} \texttt{SEP}^k=\texttt{SEP}$. 
Hence, for $\C=\texttt{SEP}$, we can relax optimizations over \eqref{maxi} by optimizing over one of the sets $\texttt{SEP}^k$ instead. Since $\texttt{SEP}^k\supset \texttt{SEP}$, the solution $f^k$ of such a semidefinite program will satisfy $f^k\geq f^\star$. Moreover, $\lim_{k\to\infty}f^k=f^\star$. For entanglement detection problems, the use of a relaxation of $\C$ in optimizations such as \eqref{induction} results in an upper bound on the maximum average game score.

To model constraints of the form  \eqref{dualita}, we similarly replace the dual of $\texttt{SEP}$ by the dual of $\texttt{SEP}^k$ in eq.\eqref{dualita}, that, as we shall show, also admits a semidefinite programming representation. Since $\texttt{SEP}^*\supset (\texttt{SEP}^k)^*$, we have that $v\id-W\in (\texttt{SEP}^k)^*$ implies $v\id-W\in \texttt{SEP}^*$. However, there might exist values of $v$ such that $v\id-W\in \texttt{SEP}^*$, but $v\id-W\not\in (\texttt{SEP}^k)^*$. Such replacements in expressions of the form \eqref{constOpt} will lead, as before, to an overestimation of the maximum average score of the game for the considered set of preparation strategies.

Let us thus work out a semidefinite representation for the set $(\texttt{SEP}^k)^*$. By duality theory \cite{sdp}, we have that any $W\in E_k^*$ must be of the form
\be
W=(C-\Pi_k C\Pi_k) +\sum_{S\in {\cal N}}  M_S^{\T_S},
\label{dual_theo}
\ee
\noindent for some positive semidefinite matrices $\{M_S\}_S$. Indeed, multiplying by $\rho_{AB_1,...,B_k}\in E_k$ and taking the trace, we find, by virtue of the defining relations \eqref{def_E} that the trace of $\rho_{AB_1,...,B_k}$ with respect to each term in the above equation is non-negative.

Multiplying on both sides of \eqref{dual_theo} by $\Pi_k$, we arrive at the equivalent condition
\be
\Pi_k W\Pi_k=\Pi_k\left(\sum_{S\in {\cal N}}  M_S^{\T_S}\right)\Pi_k.
\ee

Now, let $V\in (\texttt{SEP}^k)^*$, and let $\rho_{AB}\in \texttt{SEP}^k$ with extension $\rho_{AB_1,...,B_k}\in E_k$. Then we have that
\be
\tr[V\rho_{AB}]=\tr\left[(V\otimes\id_B^{\otimes k-1})\rho_{AB_1,...,B_2}\right]\geq 0.
\ee
\noindent Since this relation must hold for all $\rho_{AB_1,...,B_k}\in E_k$, it follows that $V\otimes\id_B^{\otimes k-1}\in E^*_k$. 
In conclusion, $V\in (\texttt{SEP}^k)^*$ iff there exist positive semidefinite matrices $\{M_S\}_S$ such that
\be
\Pi_k(V\otimes\id_B^{\otimes k-1})\Pi_k=\Pi_k\left(\sum_{S\in {\cal N}}  M_S^{\T_S}\right)\Pi_k.
\label{charac_dual}
\ee

\noindent This constraint obviously admits a semidefinite programming representation.

For $\mbox{dim}(\H_A)\mbox{dim}(\H_B)\leq 6$, $\texttt{SEP}^1=\texttt{SEP}$ \cite{horo}. In such cases, we have by eq.\eqref{charac_dual}, that
\be
\texttt{SEP}^*=\{V:V=V_0+V_1^{\T_B}, V_0,V_1\geq 0\}.
\ee

\section{Maxwell demon games for entanglement detection}
\label{app:maxwellapplications}

Here we provide the technical details regarding the applications of Maxwell demon games to entanglement certification. We consider honest players with i.i.d.\ strategies ${\cal E}=\{\rho^{\otimes n}: \rho\in E\}$, and we are interested in the worst-case errors $\max_{\rho \in E} e_{II}(M, \rho^{\otimes n})$.

In each round $k$, Alice and Bob must choose the indices $x_k, y_k$ of the measurements that they will conduct on their respective subsystems. That is, in round $k$ Alice (Bob) will conduct the measurement $\{A^{(k)}_{a|x}:a\}$ ($\{B^{(k)}_{b|y}:b\}$), with outcome $a_k$ ($b_k$). To model their (classical) decision process, we will introduce the variables $\{P(x_1, y_1,x_2,y_2,...,x_n,y_n,\gamma|a_1,b_1,...,a_n,b_n)\}$.

$P(x_1, y_1,x_2,y_2,...,x_n,y_n,\gamma|a_1,b_1,...,a_n,b_n)$  will satisfy some linear restrictions related to the no-signalling to the past condition, whose exact expression depends on how Alice and Bob conduct their measurements in each round. If, in each round, Alice and Bob make use of 1-way LOCC measurements from Alice to Bob (measurement class $\M_2$), then $P$ will satisfy the constraints
	\begin{align}
	&\sum_{\gamma}P(x_1, y_1,...,x_n,y_n,\gamma|a_1,b_1,...,a_{n},b_{n})=P(x_1, y_1,x_2,y_2,...,x_n,y_n|a_1,b_1,...,a_{n-1},b_{n-1},a_n),\nonumber\\
	&\sum_{x_k,...,x_n,y_k...y_n}P(x_1, y_1,...,x_n,y_n|a_1,b_1,...,a_{n-1},b_{n-1},a_n)=P(x_1, y_1,x_2,y_2,...,x_{k-1},y_{k-1}|a_1,b_1,...,a_{k-2},b_{k-2},a_{k-1}),\nonumber\\
	&\sum_{x_{k+1}...,x_n,y_{k},...,y_n}P(x_1, y_1,...,x_n,y_n|a_1,b_1,...,a_{n-1},b_{n-1},a_n)=P(x_1, y_1,x_2,y_2,...,x_{k}|a_1,b_1,...,a_{k-1},b_{k-1}).
	\end{align}
	If, on the contrary, Alice and Bob use local measurements in each round (measurement class $\M_3$), then the constraints on $P$ will be
	\begin{align}
	&\sum_{\gamma}P(x_1, y_1,...,x_n,y_n,\gamma|a_1,b_1,...,a_{n},b_{n})=P(x_1, y_1,x_2,y_2,...,x_n,y_n|a_1,b_1,...,a_{n-1},b_{n-1}),\nonumber\\
	&\sum_{x_k,...,x_n,y_k...y_n}P(x_1, y_1,...,x_n,y_n|a_1,b_1,...,a_{n-1},b_{n-1})=P(x_1, y_1,x_2,y_2,...,x_{k-1},y_{k-1}|a_1,b_1,...,a_{k-2}, b_{k-2}).
	\end{align}

As explained in Results, constraints \eqref{initial1}, \eqref{initial2}, \eqref{EW} require minor modifications, to take into account that, in each round, the set of effective measurements of Alice and Bob is not finite (although the set of local measurements of either party is). More specifically, we define $s_k = (x_1,y_1,a_1,b_1,...,x_{k-1},y_{k-1},a_{k-1},b_{k-1})$ to be the game configuration at the beginning of round $k$. To enforce eq. (\ref{upper_score}) for $\C$-constrained strategies (where, in this case, $\C$ denotes the set of separable quantum states), we introduce the following relations:
	\begin{align}
	&\xi_{s_{n+1}}=\sum_{\score\in \G}\score P(x_1,...,y_n,\score|a_0,b_0,...,a_n, b_n),\\
	&\nu^{(n)}_{s_n}\id-\sum_{a_{n},b_n,x_{n}, y_n}\xi_{s_{n+1}}A^{(n)}_{a_{n}|x_{n}}\otimes B^{(n)}_{b_{n}|y_{n}}\in \C^*,\nonumber\\
	&\nu^{(k)}_{s_k}\id-\sum_{a_{k},b_k,x_{k}, y_k}\nu^{(k+1)}_{s_{k+1}}A^{(k+1)}_{a_{k+1}|x_{k+1}}\otimes B^{(k+1)}_{b_{k+1}|y_{k+1}}\in \C^*.
	\end{align}

\section{The dual of $\epsilon$-balls of quantum states}
\label{sec:tracedistance}
\label{app:approx}
To enforce relation (\ref{upper_score}) when $\S$ corresponds to the set of strategies achievable by a player constrained to prepare states within an $\epsilon$-ball around a quantum state $\rho$, we need to find the dual of the set of states 
\be 
\mathcal{C}(\rho, \epsilon) = \{ \rho' : \rho' \geq 0, \ \tr(\rho')=1, \ \| \rho'-\rho \|_1  \leq \epsilon \}. \nonumber
\ee

For this purpose, let us consider the optimisation problem 
\begin{align}
&\min_{\rho'} \tr[M \rho']\nonumber\\
\mbox{s.t. }&\rho' \in \mathcal{C}(\rho, \epsilon) , 
\end{align}
which has a non-negative solution if and only if $M \in \mathcal{C}^*(\rho, \epsilon)$. This problem can be written as
\begin{align}
&\min_{\rho', Z} \tr[M \rho']\nonumber\\
\mbox{s.t. } &\rho'\geq 0,\tr(\rho')=1,\nonumber\\
& Z + \rho' -\rho \geq 0, \nonumber \\
& Z - \rho' +\rho \geq 0, \nonumber \\
& \tr[Z] =\epsilon. 
\label{primaltrace}
\end{align}
Now, note that the dual to this semi-definite program is
\begin{align}
&\max_{A, \mu, \lambda} 2\tr[A\rho] - \lambda  (1+\epsilon)-\mu \nonumber\\
\mbox{s.t. }& A \geq 0, \nonumber \\
& \lambda \id - A \geq 0, \nonumber \\
& (\mu+\lambda)\id+M-2A \geq 0,
\label{dualtrace}
\end{align}
and that the two problems are strongly dual. Thus, \eqref{dualtrace} has a non-negative solution if and only if \eqref{primaltrace} does. This implies that 
\be 
\mathcal{C}^*(\rho, \epsilon) = \{M:\exists \mu, \ \lambda\in\R, \ A\geq 0,\mbox{ s.t. } \lambda\id-A\geq 0, \ (\mu+\lambda)\id+M-2A\geq 0, \ 2\tr(A\rho)-\lambda(1+\epsilon)-\mu\geq 0\}.
\ee

\section{Round-by-round optimization of preparation games}
\label{app:implementationfincorr}

In the following, we illustrate how to optimize the POVMs of an individual round of a preparation game. This is the main subroutine in the heuristic presented in Box~2. We then describe the application of this one-round optimisation to a particular problem: entanglement detection of finitely correlated states. We further illustrate the efficiency of our coordinate-descent-based heuristic, as presented in Box~2, in this example.

To optimise over a single game round, notice that eq.~\eqref{induction} implies the conditions
\be
\mu^{(n)}_s\id- \sum_{\bar{s}\in\bar{S}}\langle g(\bar{s})\rangle M^{(n)}_{\bar{s}|s},\; \  \mu^{(k)}_s\id-\sum_{s'}M^{(k)}_{s'|s}\mu^{(k+1)}_{s'}\in \C^*.
\ee
Optimizations over $\{M^{(k)}_{s_{k+1}|s_k}:s_k,s_{k+1}\}$ under a constraint of the form \eqref{upper_score} can thus be achieved via the following convex optimization scheme: first, compute $\{\mu^{(j)}_s:j>k\}$ by induction via eq.\eqref{induction}. Next, impose the constraints
\begin{align}
	&\mu^{(1)}_{\emptyset}\leq v,\nonumber\\
	&\mu^{(j)}_s\id-\sum_{s'}M^{(j)}_{s'|s}\mu^{(j+1)}_{s'}\in \C^* , \mbox{ for } j =1,...,k.\nonumber\\
	&\{M^{(k)}_{s_{k+1}|s_k}:s_{k+1}\}\subset \M, \mbox{ for } s_k\in S_k.
	\label{constOpt}
\end{align}
Note that, in the second constraint of eq.\eqref{constOpt}, either $M^{(j)}_{s'|s}$ or $\mu^{(j+1)}_{s'}$ is an optimization variable, but not both. This means that all the above are indeed convex constraints.

Remarkably, expressing condition \eqref{upper_score} for i.i.d., finitely-correlated and $\C$-constrained strategies requires adding $O\left(\sum_{j\leq k}|S_j|\right)$ new optimization variables, related to the original ones through a set of $O\left(\sum_{j\leq k}|S_j|\right)$ constraints, all of which can be calculated with $O\left(\sum_{j}|S_j||S_{j+1}|\right)$ operations. As long as the number of game configurations is not excessive, one can therefore carry these optimizations out for games with very large $n$.

We are now ready to test the practical performance of the heuristic described in Box $2$. To this aim, consider the following entanglement detection scenario:  an honest player is attempting to prepare the maximally entangled state $\ket{\psi_{\frac{\pi}{4}}}=\frac{1}{\sqrt{2}}\left(\ket{00}+\ket{11}\right)$, but, before being transmitted, the state interacts with the local environment $\rho_A$ for a brief amount of time $\tau$. Specifically, we take the environment to be a $d_A$-dimensional quantum system that interacts with the desired state through the Hamiltonian \be
H_I = a_A^\dagger\otimes\left(\id \otimes \ket{0}\bra{1} + \ket{0}\bra{1} \otimes \id\right) + a_A\otimes\left(\id \otimes \ket{1}\bra{0} + \ket{1}\bra{0} \otimes \id\right),
\label{eq:inthamiltonian}
\ee
where $a_A^\dagger$ and $a_A$ are raising and lowering operators acting on the environmental system, respectively. We let the environment evolve only when it interacts with each new copy of $\ket{\psi_{\frac{\pi}{4}}}$. By means of global bipartite measurements $\M_1$, we wish to detect the entanglement of the states prepared by the honest player. Our goal is thus to devise adaptive measurement protocols that detect the entanglement of a family of finitely correlated strategies of fixed interaction map, but with an unknown initial environment state. 

When the player is following such a finitely correlated strategy, optimizing the $k^{th}$ round measurements amounts to solving the following semi definite program:
\begin{align}
	&\min_{M^{(k)},\{\mu^{(j)}:j\leq k\}} e_{II} \nonumber\\
	\mbox{s.t. }
	& \{ M^{(k)}_{s'|s} \}_{s'} \subset \M_1, \mbox{ for } s\in S_k \nonumber\\
	& \mu^{(1)}_\emptyset=e_I, \nonumber\\
	& \mu^{(j)}_s\id-\sum_{s'}M^{(j)}_{s'|s}\mu^{(j+1)}_{s'}\in \C^* ,\mbox{ for }j=1,...,k, \nonumber\\
	& \Omega(M^{(k)}) - (1-e_{II})\id \geq 0,
	\label{n-shot-boundedconfig}
\end{align}
\noindent where $M^{(j)}$ ($\mu^{(j)}$) stands for $\{M^{(j)}_{s'|s}:s'\in S_{j+1}, s\in S_j\}$ ($\{\mu^{(j)}_s:s\in S_j\}$) and $\Omega(M^{(k)})$ is defined according to~\eqref{recursive_omega}. The quantities $\{\mu^{(j)}:j>k\}$ do not depend on $M^{(k)}$, and hence can be computed via eq.~\eqref{induction} before running the optimization. 

\begin{figure}
	\begin{minipage}[t]{0.5\linewidth}
		\includegraphics[width=1\textwidth]{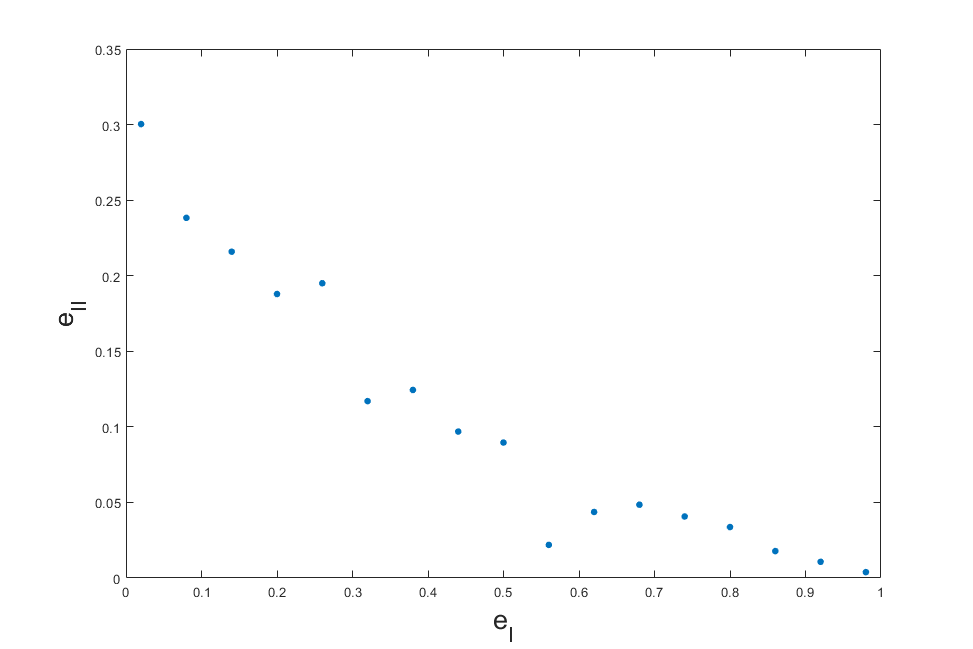}
	\end{minipage}
	\caption{Error trade-off curve for a finitely correlated scenario with bounded configuration space. The results are obtained with a $10$-dimensional unknown environment that interacts with a maximally entangled state for $\tau=0.1$ according to the Hamiltonian \eqref{eq:inthamiltonian}. There were 20 measurement rounds ($n=20$), and in each of the first $19$ rounds a $6$-outcome measurement was performed, with the option of outputting $0$ available as one of the outcomes of each measurement. These results were obtained through the method outlined in the main text. For each value of $e_I$, we plot the minimum $e_{II}$ achieved in $10$ runs (each time with a different random initialization of the measurements). Each run has been optimized until convergence was achieved. Although the type-II errors obtained are reasonably small, the curve presents large discontinuities and, in fact, is not even decreasing. Presumably, for many of the values of $e_I$, the initial (random) measurement scheme fed into the algorithm led the latter to a local minimum. This explains, e.g., the sudden drop of the type-II error after $e_I=0.5$.}
	\label{fig:fincorr}
\end{figure}

We consider a configuration space where $|S_k|=m$ for all $k=2,3,\ldots,n$, and $S_{n+1}
=\{0, 1\}$. In other words, the first $n-1$ measurements are carried out with $m$-outcome POVMs, and the last measurement is dichotomic. Furthermore, in each round, we include the possibility of terminating the game early and simply outputting $0$ (i.e., $0\in S_k$). This models a scenario where the referee is convinced early that they will not be able to confidently certify the states to be entangled. Applying the coordinate-descent heuristic in Box 2 for different values of $e_{I}$, we arrive at the plot shown in Figure \ref{fig:fincorr}. 

\end{document}